\documentclass[11pt]{article}

\usepackage[margin=.75in]{geometry}
\usepackage{datetime}
\usepackage{amsmath, amsthm, amssymb,fancyhdr,mathrsfs,  enumerate}
\usepackage{latexsym}
\usepackage[round, authoryear, comma, sort&compress]{natbib}
\usepackage{hyperref}
\usepackage[utf8]{inputenc}
\usepackage{graphicx,float}
\usepackage[export]{adjustbox}

\newtheorem{lemma}{Lemma}

\newcommand{\bpsi}{{\bf\Psi}}
\newcommand{\bb}{{\bf\beta}}
\newcommand{\bt}{{\bf\theta}}
\newcommand{\eps}{{\epsilon}}

\newcommand{\bl}{{\bf \Lambda}}
\newcommand{\ba}{{\bf \alpha}}

\newcommand{\E}{\mathbb{E}}
\newcommand{\I}{\mathbb{I}}
\newcommand{\R}{\mathbb{R}}

\newcommand{\widesim}[2][1.5]{
  \mathrel{\overset{#2}{\scalebox{#1}[1]{$\sim$}}}
}

\begin{document}

\begin{center}
      \Large{\bf {Unsupervised Liu-type Shrinkage Estimators for Mixture of Regression Models}}
\end{center}
\begin{center}
 \noindent{{\sc Elsayed Ghanem$^{\dagger,\ddagger}$},  
 {\sc Armin Hatefi$^{\dagger,}$\footnote{Corresponding author:\\
 Email: ahatefi@mun.ca and Tel: +1 (709) 864-8416}}
  and {\sc Hamid Usefi$^{\dagger}$}}

\vspace{0.5cm}
\noindent{\footnotesize{\it $^{\dagger}$Department of Mathematics and Statistics, Memorial University of Newfoundland, St. John's, NL, Canada.}} \\
\noindent{\footnotesize{\it $^{\ddagger}$Faculty of Science, Alexandria University, Arab Republic of Egypt.}}
\end{center}

\begin{center} {\small \bf Abstract}:
\end{center}
{\small
In many applications (e.g., medical studies), the population of interest (e.g., disease status) 
comprises heterogeneous subpopulations. The mixture of probabilistic regression models is one of
 the most common techniques to incorporate the information of covariates into 
learning of the population heterogeneity. Despite its flexibility, the model may lead to unreliable
 estimates in the presence of multicollinearity problem. In this paper, we develop Liu-type shrinkage methods through an
  unsupervised learning approach to estimate the model coefficients in multicollinearity. The 
  performance of the developed methods is evaluated via classification and stochastic versions 
  of EM algorithms. The numerical studies show that the proposed methods outperform their Ridge 
  and maximum likelihood counterparts. Finally, the developed methods are applied to analyze 
  the bone mineral data of women aged 50 and older.     
}


\noindent {\bf Keywords: } Multicollinearity, Maximum likelihood, Ridge penalty, Liu-type penalty
Mixture models, EM algorithm, Bone mineral data.

\section{ Introduction } \label{sec:intro}
As a bone metabolic disease, osteoporosis is characterized when the mineral density of bone tissues 
decreases significantly, leading to various major health problems such as skeletal fragility 
and osteoporotic fractures. These can occur in different body areas, including the hip, spine and femur \citep{cummings1995risk,neuburger2015impact}. Bone mineral density (BMD) is considered one of the most
 reliable predictors in determining osteoporosis status \citep{world1994assessment}. For example, 
 approximately every 1 out of 3 women and 1 out of 5 men aged 50 and older have osteoporosis and its 
 fractures \citep{melton1998bone}. 
Osteoporosis typically occurs without any significant symptoms. 
In South Korea, for example, almost 75\% of patients are unaware of their osteoporosis problem
 \citep{lim2016comparison}. BMD score of an individual improves until age 30 and declines as the individual ages. 
As the aged population is growing, it is essential to study osteoporosis to plan the well-being
 and life quality of the aged groups of the communities. 

BMD values are measured through dual-energy X-ray absorptiometry imaging and require 
a costly and time-consuming procedure. Unlike BMD scores, clinicians have access to various easily 
attainable explanatory variables about patients, including BMI, weight, age and test results from 
previous years. Linear regression models are well-known statistical tools to investigate the 
impact of a set of covariates (e.g., patients' characteristics) on a response variable (e.g., BMD measurement). 
The least squares (LS) method is a common method for estimating the regression model coefficients; 
however, the LS method can lead to extremely unreliable and misleading results in
multicollinearity when the covariates are linearly dependent. As a shrinkage method, 
Ridge regression is a common solution to the multicollinearity. A small ridge parameter 
is not enough to handle the ill-conditioned design matrix when the problem is severe. On the other hand, 
large values of the ridge parameter render more biases in the estimation.  \cite{liu2003using} 
proposed the Liu-type (LT) shrinkage method to deal with the challenge and address the
multicollinearity. \cite{duran2012difference} extended the LT
method to the semi-parametric regressions. \cite{arashi2014improved} proposed the Stein-rule LT
estimators for elliptical regression models. \cite{pearce2021multiple} 
used properties of rank-based samples to improve the LT shrinkage estimators for linear and logistic regressions.

Finite mixture models (FMMs) are probabilistic model-based tools to analyze heterogeneous populations. 
The maximum likelihood (ML) method is one of the most popular techniques for estimating the parameters
 of the FMMs. A part of the popularity comes from the Expectation-Maximization (EM) algorithm 
 \citep{dempster1977maximum} that enables computing the ML estimates of the FMMs. The EM algorithm 
 decomposes the ML estimation procedure 
into E- and M-steps and iteratively estimates the parameters of the FMMs. Stochastic EM (SEM) algorithm
 \citep{celeux1985sem} was developed as a stochastic teacher of the probabilistic EM algorithm for 
 mixture models. Also, \cite{celeux1992classification} proposed a classification EM (CEM) algorithm 
 in which a classification step is implemented in the iteration for maximization based on classified 
 data. The mixture of regression models \citep{quandt1978estimating} incorporates the properties of the FMMs 
into linear regression models. \cite{jones1992fitting,hawkins2001determining} 
studied the maximum likelihood and EM algorithm to fit a mixture of regression models. \cite{faria2010fitting}
 investigated various EM algorithms for the mixture of linear regression models. FMMs have found many applications
  in the core of statistical sciences, such as times 
 series data \citep{zhang2006mixture}, sampling methods and censored data \citep{hatefi2018,wedel1998mixture,hatefi2015mixture}. For
more details about the theory and applications of FMMS, see \cite{mclachlan2019finite} and references therein.

Despite the flexibility of the mixture of linear regressions, the ML estimates of the mixture coefficients become unreliable 
in multicollinearity. Recently, \cite{elsayed_log} proposed shrinkage estimation methods for the mixture logistic
 regressions. In the manuscript, we develop the Liu-type shrinkage estimators for the mixture of linear 
 regression models where the ridge EM algorithms may not be able to address the ill-conditioned 
 component design metrics. Through extensive numerical studies, we show that the classification and 
 stochastic EM algorithms of the LT shrinkage method outperform their counterparts and provide 
 more reliable coefficient estimates for component regressions. We finally apply the developed methods 
 to analyze the bone disorder status of women aged 50 and older. 
  
  The outline of the paper is as follows: Section \ref{sec:meth} develops the shrinkage methods and their EM algorithms. 
  Section \ref{sec:sim} evaluates the performance of the  methods by simulation studies. The methods are applied 
  for analysis of bone data in Section \ref{sec:real}. Section \ref{sec:sum} finally presents the summary and concluding remarks. 
    
\section{Statistical Methods}\label{sec:meth}

Let ${\bf x}_i^\top =(x_{i,1},\ldots,x_{i,p})$ be the vector of $p$ explanatory variables for the $i$-th
 subject in a random sample of size $n$.
Let ${\bf y}=(y_1,\ldots,y_n)$ and  ${\bf X}=({\bf x}_1^\top,\ldots,{\bf x}_n^\top)^\top$ denote the response
 vector and ($n \times p$) design matrix with $\text{rank}({\bf X})=p < n$.
The Regression model  $y_i = {\bf x}_i^\top {\bb} + \epsilon_i$ for $i=1,\ldots, n$ is one of the most common statistical methods to 
study the relationship between response variable and the set of the explanatory variables. 

The mixture of regression models is a generalization of the regression model when the underlying population 
comprises of heterogeneous subpopulations. The mixture of regression models is defined by 
\begin{align} \label{mix_reg}
 y_i = \left\{
 \begin{array}{lc}
 {\bf x}_i^\top {\bb}_1 + \epsilon_{i1} & \text{with probability $\pi_1$}\\
 \vdots \\
 {\bf x}_i^\top {\bb}_J + \epsilon_{iJ} & \text{with probability $\pi_J$}\\
 \end{array} \right.
 \end{align}  
where ${\bf \beta}_j =(\beta_{j,1},\ldots,\beta_{J,p})$ represents the coefficients of $p$ predictors in the $j$-th component 
regression for $j=1,\ldots,J$ and  ${\pi}=(\pi_1,\ldots,\pi_J)$ denotes the vector of mixing proportions with $\sum_{j=1}^{M} \pi_j =1$ and $0 < \pi_j < 1$. 
Also $\epsilon_{ij}$ are independent normal random errors from each component  of the mixture; that is $\epsilon_{ij} \overset{iid}{\sim} N(0,\sigma_j^2)$.
Although we assume that the number of components $J$ in the mixture model \eqref{mix_reg} is known throughout the paper, the component  memberships of the observations are unknown 
and should be estimated in an unsupervised approach. 
Let ${\bb}=(\bb_1,\ldots,\bb_J)$. Let $\bt_j=(\bb_j,\sigma_j^2)$ denote the parameters of the $j$-th component. Thus, we  
represent the vector of all unknown parameters of the mixture \eqref{mix_reg} with $\bpsi=({\bf\pi},\bt_1,\ldots,\bt_J)$.

From regression model \eqref{mix_reg}, the log-likelihood function of $\bpsi$ can be written as
\begin{align} \label{l_ml}
\ell(\bpsi) = 
\sum_{i=1}^{n} \log \left(
\sum_{j=1}^{J} \pi_j \phi_j({\bf x}_i^\top {\bb}_j, \sigma_j^2)
\right),
\end{align}
where $\phi_j({\bf x}_i^\top {\bb}_j, \sigma_j^2)$ represents the pdf of the univariate normal 
distribution with mean ${\bf x}_i^\top {\bb}_j$ and variance $\sigma_j^2$.
We have to maximize \eqref{l_ml} to obtain the ML estimate of $\bpsi$. 
The gradient of \eqref{l_ml} is not tractable with respect to component parameters $\bt_j, j=1,\ldots,J$. 
We view $({\bf X},{\bf y})$ as incomplete data and apply the expectation-maximization (EM) algorithm of \citep{dempster1977maximum} using a complete data  to find $\widehat{\bpsi}_{ML}$.
For each subject $({\bf x}_i,y_i)$, we introduce latent variables ${\bf Z}_i=(Z_{i1},\ldots,Z_{iJ})$ for $i=1,\ldots,n$ as 
  \begin{align*}
 Z_{ij} = \left\{
 \begin{array}{lc}
 1 & \text{if the $i$-th subject comes from the $j$-th component,}\\
 0 & o.w.,
 \end{array} \right.
 \end{align*}  
where ${\bf Z}_i \widesim{iid} \text{Multi}(1,\pi_1,\ldots,\pi_M)$. 
From the marginal distribution of latent variables, the conditional distribution of ${\bf Z}_i|y_i$  is given by 
 \begin{align}\label{z|y}
f({\bf z}_i|y_i) = \prod_{j=1}^{J} 
\left\{
\frac{\pi_j \phi_j({\bf x}_i^\top {\bb}_j, \sigma_j^2)}{ \sum_{j=1}^{J} \pi_j \phi_j({\bf x}_i^\top {\bb}_j, \sigma_j^2)}
\right\}^{z_{ij}}.
\end{align}
From above, it is easily seen that   ${\bf Z}_i|y_i \widesim{iid} \text{Multi}(1, \tau_{i1}(\bpsi),\ldots,\tau_{iJ}(\bpsi))$ where
 \begin{align}\label{tau_ml}
 \tau_{ij}(\bpsi) = \frac{\pi_j \phi_j({\bf x}_i^\top {\bb}_j, \sigma_j^2)}{\sum_{j=1}^{J} \pi_j \phi_j({\bf x}_i^\top {\bb}_j, \sigma_j^2)}.
  \end{align} 
  Let $({\bf X},{\bf y},{\bf Z})$ denote the complete data. Thus,   the complete log-likelihood function of $\bpsi$ is given by 
 \begin{align}\label{ll_comp_ml}
 \ell_c(\bpsi) = \sum_{i=1}^{n} \sum_{j=1}^{J} z_{ij} \log(\pi_j)
 + \sum_{i=1}^{n} \sum_{j=1}^{J} z_{ij} \log \{\phi_j({\bf x}_i^\top {\bb}_j, \sigma_j^2)\}.
\end{align}

\subsection{ML Estimation Method}\label{sub:ml}
The EM algorithm is a standard method to find the ML estimates of the mixture model parameters.
 The EM algorithm employs the latent variables on top of the observed data and decomposes the estimation procedure into an iterative expectation (E) and maximization (M) steps.
As an iterative method, EM algorithm begins with an initial value. Let $\bpsi^{(0)}=({\bf \pi}^{(0)}, \bt_1^{(0)},\ldots,\bt_J^{(0)})$ and $\bpsi^{(r)}$ 
denote the initial vector and the estimate in the $r$-th iteration of the EM algorithm, respectively. 

On the $(r+1)$-th iteration, we require to compute the conditional expectation of  \eqref{ll_comp_ml} in the E-step. The  $Q(\bpsi,\bpsi^{(r)})$ replaces the latent variables by their conditional expectations as
\[
{\bf Q}(\bpsi,\bpsi^{(r)})  = {\bf Q}_1({\bf \pi},\bpsi^{(r)})  + {\bf Q}_2({\bt},\bpsi^{(r)}), 
\] 
where 
\begin{align} \label{Q1_ml}
{\bf Q}_1({\bf \pi},\bpsi^{(r)}) = \sum_{i=1}^{n} \sum_{j=1}^{J} \tau_{ij}(\bpsi^{(r)}) \log(\pi_j),
\end{align}
and 
\begin{align} \label{Q2-ml}
{\bf Q}_2({\bt},\bpsi^{(r)}) = \sum_{i=1}^{n} \sum_{j=1}^{J} \tau_{ij}(\bpsi^{(r)}) 
\log \left\{ \phi_j({\bf x}_i^\top {\bb}_j, \sigma_j^2)
\right\},
\end{align}
where $\tau_{ij}(\bpsi^{(r)})$ is obtained by \eqref{tau_ml}. In the M-step, we maximize ${\bf Q}(\bpsi,\bpsi^{(r)})$ with 
respect to ${\bf \pi}$ and ${\bt}$ to update $\bpsi^{(r+1)}$.
One can update ${\bf \pi}^{(r+1)}$ by maximizing ${\bf Q}_1({\bf \pi},\bpsi^{(r)})$  subject to $\sum_{j=1}^{J} \pi_j=1$ as follows
\begin{align}\label{pihat_em_ml}
{\widehat \pi}_j^{(r+1)} = \sum_{i=1}^{n}  \tau_{ij}(\bpsi^{(r)}) /n;  ~~~~ j=1,\ldots,J-1.
\end{align}

The maximization of ${\bf Q}_2({\bt},\bpsi^{(r)})$ can be reformulated by the weighted least squares (WLS) method as 
\begin{align}\label{wls_em_ml}
{\widehat \bb}_j^{(r+1)} = \underset{\bb_j}{\arg \min} ~ ({\bf y}-{\bf X} \bb)^\top {\bf W}_j ({\bf y}-{\bf X} \bb) / n,
\end{align}
where ${\bf W}_j$ is $n \times n$ diagonal matrix with diagonal elements $(\tau_{ij}(\bpsi^{(r)}),\ldots, \tau_{nj}(\bpsi^{(r)}))$ 
for all $j=1,\ldots, J$. One can easily update ${\widehat \bb}_j^{(r+1)}$ as the solution to \eqref{wls_em_ml} by
\begin{align}\label{bj_em_ml}
{\widehat \bb}_j^{(r+1)} = \left({\bf X}^\top {\bf W}_j {\bf X}\right)^{-1} {\bf X}^\top {\bf W}_j {\bf y}, ~~~ j=1,\ldots,J.
\end{align}
From \eqref{wls_em_ml} and following \citep{faria2010fitting}, we then update ${\widehat\sigma}_j^{2(r+1)}$ as follows
\begin{align}\label{sigj_em_ml}
{\widehat\sigma}_j^{2(r+1)} = \frac{({\bf y}-{\bf X} \widehat{\bb}^{(r+1)})^\top {\bf W}^{(r)}_j ({\bf y}-{\bf X} \widehat{\bb}^{(r+1)}) }{\sum_{i=1}^{n} \tau_{ij}(\bpsi^{(r)})}, ~~~ j=1,\ldots,J.
\end{align}

To find ${\widehat\bpsi}_{ML}$, we iteratively alternate the E- and M- steps of the EM algorithm until the stopping 
criterion $|\ell(\bpsi^{(r+1)})-\ell(\bpsi^{(r)})|$ becomes negligible. 

 {\bf Classification EM Algorithm}: In the above EM algorithm, we use information from all observations (as membership probabilities) 
 in each iteration to estimate  the parameters of the mixture model. 
 Following \cite{celeux1992classification}, we shall employ the classification version of EM algorithm (CEM) to estimate $\bpsi$.
 The CEM algorithm incorporates a classification (C) step between E and M steps so that the component parameters of the mixture are updated in M-step using the classified complete data log-likelihood function.
 
 The E-step here is identical to the E-step of the EM algorithm. In C-step, 
 the observations are then assigned to $J$ mutually exclusive partitions corresponding to the $J$ components of mixture model \eqref{mix_reg}.\
 Let ${\bf P}^{(r+1)}=(P_1^{(r+1)},\ldots,P_J^{(r+1)})$ denote the partition in the $(r+1)$-th iteration. 
 Each subject $({\bf x}_i,y_i)$ is assigned  to partition $P_h^{(r+1)}$ when
 \[
 \tau_{ih}(\bpsi^{(r)}) = \underset{j}{\arg\min} ~ \tau_{ij}(\bpsi^{(r)}).
 \]
 Note when the maximum weight is not unique, the tie is broken at random. Also the CEM algorithm is stopped and $\bpsi^{(r)}$ is returned when a partition becomes empty or has only one observation. 
  
  In the M-step, we maximize conditional expectation function using  ${\bf P}^{(r+1)}$. From \eqref{Q1_ml}, the mixing proportion is updated by 
  \begin{align}\label{pihat_cem_ml}
{\widehat \pi}_j^{(r+1)} = n_j /n;  ~~~~ j=1,\ldots,J,
\end{align}
  where $n_j$ is the number of observations allocated to partition $P_j^{(r+1)}$. Applying the WLS to each partition $P_j^{(r+1)}$, 
  we can update the parameters of the $j$-th component ($j=1,\ldots, J$) by
   \begin{align}\label{bj_cem_ml}
{\widehat \bb}_j^{(r+1)} = \left({\bf X}_j^\top {\bf W}_j {\bf X}_j\right)^{-1} {\bf X}_j^\top {\bf W}_j {\bf y}_j,
\end{align}
\begin{align}\label{sigj_cem_ml}
{\widehat\sigma}_j^{2(r+1)} = \frac{({\bf y}_j-{\bf X}_j \widehat{\bb}^{(r+1)})^\top {\bf W}^{(r)}_j ({\bf y}_j-{\bf X}_j \widehat{\bb}^{(r+1)}) }{\sum_{i=1}^{n} \tau_{ij}(\bpsi^{(r)})}, \end{align}
  where ${\bf X}_j$  and ${\bf y}_j$ represent, respectively, $(n_j \times p)$ design matrix and vector of responses corresponding to  $P_j^{(r+1)}$. 
  Also ${\bf W}^{(r)}_j$ is the diagonal weight matrix of size $n_j$ with entries  $(\tau_{ij}(\bpsi^{(r)}),\ldots, \tau_{n_j,j}(\bpsi^{(r)}))$.
  Finally, we alternate repeatedly the E-, C- and M- steps until  $|\ell(\bpsi^{(r+1)})-\ell(\bpsi^{(r)})|< \eps$. 
  
  {\bf Stochastic EM Algorithm}:
  One can also apply the stochastic version of the EM algorithm \citep{celeux1985sem} to fit the mixture of regression models. 
  The stochastic EM  (SEM) algorithm implements a stochastic version of the C-step between E- and M-steps. 
  Although the E- and M-steps are identical to the CEM algorithm, the SEM simulates  the component membership of each 
  observation using conditional distribution of the latent variable given incomplete data. 
  On the $(r+1)$-th iteration, the S-step simulates a random allocation for each observation using one draw out of $J$ components as 
  \[
  {\bf Z}_i^* =(Z_{i1}^*,\ldots,Z_{iJ}^*) \widesim{iid} \text{Multi}(1, \tau_{i1}(\bpsi^{(r)}),\ldots,\tau_{iJ}(\bpsi^{(r)})) ~~ (i=1,\ldots,n).
  \]
  Then the $({\bf x}_i,y_i)$ is classified to partition $P_j^{(r+1)}$ if $Z_{ij}^*=1; i=1,\ldots,j=1,\ldots,J$. 
  Using the stochastic partitions, we update the mixture parameters from \eqref{pihat_cem_ml}, \eqref{bj_cem_ml} and \eqref{sigj_cem_ml} in M-step.  
  From \citep{celeux1985sem,faria2010fitting}, point-wise convergence in the SEM algorithm is not guaranteed. The algorithm resembles a Markov chain where at stationary state fluctuates  around the ML estimate. Hence, we alternate the E-, S- and M-steps until either the criterion $|\ell(\bpsi^{(r+1)})-\ell(\bpsi^{(r)})|< \epsilon$ satisfies or the chain reaches a pre-specified  maximum number of iterations that is fixed for all the algorithms for fair comparison. 
  
\subsection{Ridge Estimation Method}\label{sub:ridge}
The ML method is a standard tool to estimate ${\bpsi}$; however, the ML estimates are 
dramatically influenced  by multicollinearity where the covariates are linearly dependent. 
Ridge method \citep{hoerl1970ridge} is one of the most common methods to encounter with the challenges of the LS method. 
The ridge estimate for \eqref{mix_reg} can be obtained as a solution to the penalized log-likelihood function  given by
 \begin{align} \label{l_ridge}
\ell^R(\bpsi) = \ell(\bpsi) - k \bb^\top \bb /2
\end{align} 
where $\ell(\bpsi)$  comes from \eqref{l_ml} and $k>0$ is the ridge parameter. 
In a similar vein to Subsection \ref{sub:ml}, for each  observation $({\bf x}_i,y_i), i=1,\ldots,n$, we first introduce $J$ dimensional latent vector ${\bf Z}_i= (Z_{i1},\ldots,Z_{iJ})$.
We then develop an EM algorithm   to maximize the complete ridge log-likelihood function and obtain ${\widehat\bpsi}_R$.
  
The E-step of the ridge EM algorithm is identical to the E-step of Subsection \ref{sub:ml}.  In the M-step, the mixing proportion are updated from \eqref{pihat_em_ml}. 
To update the coefficients  of the component regressions, we require to maximize ${\bf Q}_2({\bt},\bpsi^{(r)})$ subject to the ridge penalty within each component as 
\[
{\bf Q}^R_2({\bt},\bpsi^{(r)}) = {\bf Q}_2({\bt},\bpsi^{(r)}) - k_j \bb_j^\top \bb_j /2,
\] 
where  ${\bf Q}_2({\bt},\bpsi^{(r)})$ is from \eqref{Q2-ml} and $k_j$ is the ridge parameter in the $j$-th component. Like ML method, one can re-write the maximization of 
${\bf Q}^R_2({\bt},\bpsi^{(r)})$  as a WLS subject to the ridge penalty by
\begin{align}\label{wls_em_ridge}
{\widehat \bb}_{R,j}^{(r+1)} = \underset{\bb_j}{\arg \min} ~ ({\bf y}-{\bf X} \bb)^\top {\bf W}_j ({\bf y}-{\bf X} \bb) + k_j \bb_j^\top \bb_j /2,
\end{align}
where ${\bf W}_j$ is $n \times n$ diagonal matrix with elements $(\tau_{ij}(\bpsi^{(r)}),\ldots, \tau_{nj}(\bpsi^{(r)}))$ obtained from \eqref{tau_ml}. 
Applying \eqref{wls_em_ridge}, ${\widehat \bb}_j^{(r+1)};j=1,\ldots,J$ is updated by
\begin{align}\label{bej_em_ridge}
{\widehat \bb}_{R,j}^{(r+1)} = \left({\bf X}^\top {\bf W}_j {\bf X} + k_j \I \right)^{-1} 
{\bf X}^\top {\bf W}_j {\bf y}.
\end{align}
\begin{lemma} \label{can_em_ridge}
Under the assumptions of mixture of regression models \eqref{mix_reg},  suppose $\lambda_{1j},\ldots,\lambda_{pj}$ and $u_{1j},\ldots,u_{pj}$ be 
eigenvalues and orthonormal eigenvectors of ${\bf X}^\top {\bf W}_j {\bf X}$ where ${\bf W}_j$ is $n \times n$ diagonal matrix with 
entries $(\tau_{ij}(\bpsi^{(r)}),\ldots, \tau_{nj}(\bpsi^{(r)}))$ under ridge EM algorithm. Let ${\bl}_j=\text{diag}(\lambda_{1j},\ldots,\lambda_{pj})$ and
 ${\bf U}_j=[u_{1j},\ldots,u_{pj}]$. Then the canonical weighted ridge estimator in each component regression is given by
 \[
 {\widehat\ba}_{R,j} = \left( {\bl}_j + k_j \I \right)^{-1} {\bl}_j^{1/2} {\bf V}_1^\top {\bf W}_j^{1/2} {\bf y},
 \]
 and
  \[
  {\widehat\bb}_{R,j} = {\bf U}_j {\widehat\ba}_{R,j},
  \]
 with ${\bf V}_1=[v_{1j},\ldots,v_{pj}]$ where $v_{1j},\ldots,v_{pj}$ are the orthonormal eigenvectors of ${\bf W}_j^{1/2} {\bf X} {\bf X}^\top {\bf W}_j^{1/2}$.
 \end{lemma}
From   \eqref{wls_em_ridge},  the variance term can be updated by
\begin{align}\label{sigj_em_ridge}
{\widehat\sigma}_{R,j}^{2(r+1)} = \frac{({\bf y}-{\bf X} \widehat{\bb}_R^{(r+1)})^\top {\bf W}^{(r)}_j ({\bf y}-{\bf X} \widehat{\bb}_R^{(r+1)}) }{\sum_{i=1}^{n} \tau_{ij}(\bpsi^{(r)})}, 
\end{align}
 where $\widehat{\bb}_R^{(r+1)}=(\widehat{\bb}_{R,1}^{(r+1)},\ldots,\widehat{\bb}_{R,J}^{(r+1)})$.
 There are various methods available in the literature for the estimation of $k_j$.  Following \cite{hoerl1975ridge,liu2003using}, we estimate the parameter by 
 $\widehat{k}_j=p {\widehat\sigma}_{ML,j}^{2}/ {\widehat \bb}_{ML,j}^\top {\widehat \bb}_{ML,j}$ where 
 ${\widehat\sigma}_{ML,j}^{2}$ and ${\widehat \bb}_{ML,j}$ are calculated from \eqref{sigj_em_ml} and \eqref{bj_em_ml}, respectively. 
 The E- and M-steps are repeatedly computed until  $|\ell^R(\bpsi^{(r+1)})-\ell^R(\bpsi^{(r)})|< \epsilon$. 
 
 {\bf Ridge CEM Algorithm}:
 One can also apply the CEM algorithm to find ${\widehat\bpsi}_R$. 
 Similar to the CEM algorithm in Subsection \ref{sub:ml}, we require to accommodate a C-step between E- and M-steps in the ridge EM algorithm. 
 Here the E-step remains the same as before. Similar to the C-step of ML method,  we classify the observations to partitions 
 ${\bf P}^{(r+1)}=(P_1^{(r+1)},\ldots,P_J^{(r+1)})$ based on the maximum probability of memberships; that is
 \[
 P_j^{(r+1)} = \{({\bf x}_i,y_i); \tau_{ij}(\bpsi^{(r)}) = \underset{h}{\arg\max} ~ \tau_{ih}(\bpsi^{(r)})\}, ~~ \forall j=1,\ldots,J.
 \] 
 Based on ${\bf P}^{(r+1)}$,  we use \eqref{pihat_cem_ml} to update the mixing proportions of the mixture. The ridge parameters are estimated similar to the ridge EM algorithm. 
 We apply \eqref{wls_em_ridge} to each partition $P_j^{(r+1)}$ and update the coefficients and variance term of each component regression  by
 \begin{align}\label{bej_cem_ridge}
{\widehat \bb}_{R,j}^{(r+1)} = \left({\bf X}_j^\top {\bf W}_j {\bf X}_j + k_j \I \right)^{-1} 
{\bf X}_j^\top {\bf W}_j {\bf y}_j,
\end{align}
\begin{align}\label{sigj_cem_ridge}
{\widehat\sigma}_{R,j}^{2(r+1)} = \frac{({\bf y}_j-{\bf X}_j \widehat{\bb}_R^{(r+1)})^\top {\bf W}^{(r)}_j ({\bf y}_j-{\bf X}_j \widehat{\bb}_R^{(r+1)}) }{\sum_{i=1}^{n} \tau_{ij}(\bpsi^{(r)})}, 
\end{align}
 where ${\bf X}_j$ is $(n_j \times p)$ design matrix and ${\bf y}_j$ is vector of responses from observations classified to  $P_j^{(r+1)}$. 
 ${\bf W}^{(r)}_j$ is the diagonal weight matrix  with entries  $\left(\tau_{ij}(\bpsi^{(r)}),\ldots, \tau_{n_j,j}(\bpsi^{(r)})\right)$ from \eqref{tau_ml}. 
  Finally, the E-, C- and M- steps under ridge estimation procedure are alternated until convergence criterion is statified. 

  {\bf Ridge SEM Algorithm}:
  Ridge estimation method can be implemented via the stochastic EM algorithm.  Like SEM of ML method, 
  the S-step determines stochastically the component membership of observations under ridge method by 
  ${\bf Z}_i^* =(Z_{i1}^*,\ldots,Z_{iJ}^*) \widesim{iid} \text{Multi}(1, \tau_{i1}(\bpsi^{(r)}),\ldots,\tau_{iJ}(\bpsi^{(r)})); i=1,\ldots,n
  $ and updates  ${\bf P}^{(r+1)}=(P_1^{(r+1)},\ldots,P_J^{(r+1)})$ such that $P_j^{(r+1)}=\{ ({\bf x}_i,y_i); Z_{ij}^*=1\}; \forall j=1,\ldots,J$. 
 Based on this stochastic partition of S-step, we update the mixture parameters by \eqref{pihat_cem_ml}, \eqref{bej_cem_ridge} and \eqref{sigj_cem_ridge}.  
  \begin{lemma} \label{can_cem_ridge}
Under the assumptions of mixture of regression models \eqref{mix_reg}, with component regression models ${\bf y}_j = {\bf X}_j \bb_j +\eps$ based on $n_j$ 
observations with $\text{rank}({\bf X}_j)=p$. 
 Suppose $\lambda_{1j},\ldots,\lambda_{pj}$ and $u_{1j},\ldots,u_{pj}$ be 
eigenvalues and orthonormal eigenvectors of ${\bf X}_j^\top {\bf W}_j {\bf X}_j$ where ${\bf W}_j$ is $n_j \times n_j$ diagonal matrix with 
entries $(\tau_{ij}(\bpsi^{(r)}),\ldots, \tau_{nj}(\bpsi^{(r)}))$ under the ridge CEM  or ridge SEM algorithm. Let ${\bl}_j=\text{diag}(\lambda_{1j},\ldots,\lambda_{pj})$ and
 ${\bf U}_j=[u_{1j},\ldots,u_{pj}]$. Then The canonical weighted ridge estimator in each component regression is given by
 \[
 {\widehat\ba}_{R,j} = \left( {\bl}_j + k_j\I\right)^{-1} {\bl}_j^{1/2} {\bf V}_1^\top {\bf W}_j^{1/2} {\bf y}, 
 \]
 and
  \[
  {\widehat\bb}_{R,j} = {\bf U}_j {\widehat\ba}_{R,j},
  \]
 with ${\bf V}_1=[v_{1j},\ldots,v_{pj}]$ where $v_{1j},\ldots,v_{pj}$ are the orthonormal eigenvectors of ${\bf W}_j^{1/2} {\bf X}_j {\bf X}_j^\top {\bf W}_j^{1/2}$.
 \end{lemma}

 Finally, the E-, S- and M-steps are iterated until the stopping rule is satisfied or the algorithm 
 reaches a pre-specified maximum number of iterations.

\subsection{Liu-type Estimation Method}\label{sub:liu}
When the design matrix is severely ill-conditioned, adding small values to the diagonal elements 
by the ridge estimator may be unable to cope with the problem. On the other side, increasing the 
ridge parameter may result in a more considerable bias in the ridge estimation method. 
\cite{liu2003using} proposed Liu-type (LT) shrinkage method for the multicollinearity in 
estimating the regression parameters. Like the ridge method, the LT method optimizes the 
estimating equation subject to the LT penalty to control the multicollinearity.  
The LT penalty is given by
 \begin{align}\label{lt_penalty}
(-\frac{d}{k^{1/2}}) {\widehat \bb} = k^{1/2} \bb + \epsilon',
\end{align}
where ${\widehat \bb}$ can be any estimator of coefficients and $d \in \R$ and $\lambda > 0$ are two 
parameters of the LT method. 
 We develop the LT shrinkage method in estimating the unknown parameters of the 
mixture model \eqref{mix_reg}. We shall find the LT estimate of $\bpsi$ by maximizing the log-likelihood
 function \eqref{l_ml} subject to the LT penalty.

 In a similar vein to Subsection \ref{sub:ml}, the penalized log-likelihood function based on 
 observed data is not tractable with respect to the component parameters. We develop the LT 
 estimation procedure through an unsupervised approach and use the EM algorithm to estimate ${\bpsi}$. 
 We again require latent 
 vectors ${\bf Z}_i=(Z_{i1},\ldots,Z_{iJ})$ to represent the component membership of the $i$-th 
 observation $({\bf x}_i,y_i);i=1,\ldots,n$. Let $({\bf X},{\bf y},{\bf Z})$ denote the complete 
 data. Then EM algorithm under the LT method proceeds as follows.

In $(r+1)$-th iteration, the E-step remains identical to the E-step of the ML method. 
The mixing proportions under the LT method are updated by \eqref{pihat_em_ml}.
We require to maximize ${\bf Q}_2({\bt},\bpsi^{(r)})$ from \eqref{Q2-ml} under the LT penalty 
within each component to estimate the component parameters.  
The LT penalized log-likelihood function can be written as a WLS constrained on the LT penalty as 
\begin{align}\label{wls_em_lt}
{\widehat \bb}_{LT,j}^{(r+1)} = \underset{\bb_j}{\arg \min} ~ ({\bf y}-{\bf X} \bb)^\top {\bf W}_j ({\bf y}-{\bf X} \bb) + 
\left[(-\frac{d_j}{k_j^{1/2}}) {\widehat \bb}_j - k_j^{1/2} \bb_j \right]^\top \left[(-\frac{d_j}{k_j^{1/2}}) {\widehat \bb}_j - k_j^{1/2} \bb_j \right],
\end{align}
where ${\widehat \bb}_j$ can be any estimate and ${\bf W}_j$ is a weight diagonal matrix with
 elements 
$(\tau_{1j}(\bpsi^{(r)}),\ldots,\tau_{nj}(\bpsi^{(r)}); j=1,\ldots, J$.  From \eqref{wls_em_lt}, the coefficients
 and variance terms in each component regression are updated by
 \begin{align}\label{bej_em_lt}
{\widehat \bb}_{LT,j}^{(r+1)} = \left({\bf X}^\top {\bf W}_j {\bf X} + k_j \I \right)^{-1} 
({\bf X}^\top {\bf W}_j {\bf y}-d_j {\widehat \bb}_j),
\end{align}
\begin{align}\label{sigj_em_lm}
{\widehat\sigma}_{LT,j}^{2(r+1)} = \frac{({\bf y}-{\bf X} \widehat{\bb}_{LT}^{(r+1)})^\top {\bf W}^{(r)}_j 
({\bf y}-{\bf X} \widehat{\bb}_{LT}^{(r+1)}) }{\sum_{i=1}^{n} \tau_{ij}(\bpsi^{(r)})}, 
\end{align}
where $\widehat{\bb}_{LT}^{(r+1)}=(\widehat{\bb}_{LT,1}^{(r+1)},\ldots,\widehat{\bb}_{LT,J}^{(r+1)})$.
The \eqref{bej_em_lt} and \eqref{sigj_em_lm} require the estimates of the LT
 parameters $(k_j,d_j)$ for each component regression. 
From \cite{liu2003using}, we can estimate $k_j$ in the $j$-th component by 
${\widehat k}_{LT,j} = \lambda_{1,j} - 100 \lambda_{p,j}/99$ where $\lambda_{1,j}$ and $\lambda_{p,j}$ 
are the maximum and the minimum 
eigenvalues of ${\bf X}^\top {\bf W}_j {\bf X}$ on the $(r+1)$-the iteration of the EM algorithm.
\begin{lemma} \label{can_em_lt}
Under the assumptions of Lemma \eqref{can_em_ridge}, the canonical LT estimator in the $j$-th component 
regression $j=1,\ldots,J$ under EM algorithm is given by 
 \[
 {\widehat\ba}_{LT,j} = \left( {\bl}_j + k_j \I\right)^{-1} ({\bl}_j^{1/2} {\bf V}_1^\top {\bf W}_j^{1/2} 
 {\bf y}- d_j {\widehat\ba}_{j}), 
 \]
 and
  \[
  {\widehat\bb}_{LT,j} = {\bf U}_j {\widehat\ba}_{LT,j},
  \]
 where ${\widehat\ba}_{j}$  is the canonical estimate of $\bb_j$ and ${\bf V}_1=[v_{1j},\ldots,v_{pj}]$ with $v_{1j},\ldots,v_{pj}$ are 
 the orthonormal eigenvectors of ${\bf W}_j^{1/2} {\bf X} {\bf X}^\top {\bf W}_j^{1/2}$.    
 \end{lemma}
 Following \cite{liu2003using} and Lemma \ref{can_em_lt}, the optimal $d_j$  can be obtained by the next lemma within 
 each component of the mixture of regression models.
\begin{lemma} \label{dj_em}
Under the assumptions of Lemma \ref{can_em_ridge} for $k_j>0$ 
\begin{itemize}
\item[i)] when ${\widehat\ba}_j = {\widehat\ba}_{ML,j}$, then
 \[
  d_j = \sum_{m=1}^{p} \left((\sigma_j^2-k_j \alpha_{mj}^2)/(\lambda_{mj}+k_j)^2\right)/\sum_{m=1}^{p}
  \left((\lambda_{mj}\alpha_{mj}^2+\sigma_j^2)/\lambda_{mj} (\lambda_{mj}+k_j)^2\right),
 \]
 \item[ii)] when ${\widehat\ba}_j = {\widehat\ba}_{R,j}$, then
 \[
  d_j = \sum_{m=1}^{p} \left(\lambda_{mj}(\sigma_j^2-k_j \alpha_{mj}^2)/(\lambda_{mj}+k_j)^3\right)/\sum_{m=1}^{p}
  \left(\lambda_{mj}(\lambda_{mj}\alpha_{mj}^2+\sigma_j^2)/ (\lambda_{mj}+k_j)^4\right),
 \]
 \end{itemize}
 minimizes the $\text{MSE}({\widehat\ba}_{LT,j})$ within each component \eqref{mix_reg}
  in the EM algorithm of the LT method. 
  \end{lemma}
Although Lemma \ref{dj_em} paves the path in estimating the optimal LT parameter $d_j$ within each 
component regression, the optimal value still depends on the unknown quantities including $\sigma_j$, $k_j$, $\ba_j$ 
and $\lambda_{m,j}$ for $m=1,\ldots,p$
 and $j=1,\ldots,J$. From Lemma \ref{dj_em}, we propose a practical approach where $d_j, j=1,\ldots,J$ 
 can be updated in the 
 $(r+1)$-the iteration of the EM algorithm by
\begin{align}\label{djhat_em}
  {\widehat d}_j = \sum_{m=1}^{p} \left( \lambda_{mj}({\widehat\sigma}_{R,j}^2-{\widehat k}_{j} {\widehat\alpha}_{R,mj}^2)/(\lambda_{mj}+
  {\widehat k}_{j})^3\right)/\sum_{m=1}^{p}\left( \lambda_{mj}(\lambda_{mj}{\widehat\alpha}_{R,mj}^2+{\widehat\sigma}_{R,j}^4)/
  \lambda_{mj} (\lambda_{mj}+{\widehat k}_{j})^2\right),
\end{align}
where ${\widehat k}_j={\widehat k}_{LT,j}$, ${\widehat\ba}_{R,j}= ({\widehat\alpha}_{R,1j},\ldots,{\widehat\alpha}_{R,pj})$
 is given by Lemma \ref{can_em_ridge} 
and $(\lambda_{1j},\ldots,\lambda_{pj})$ are eigenvalues of ${\bf X}_j^\top {\bf W}_j {\bf X}_j$ 
with ${\widehat\sigma}_{R,j}^2$ from \eqref{sigj_em_ridge}.
The E- and M-steps are alternated until the stopping criterion is satisfied. 
As the parameters $k_j$ and $d_j$ are updated in each iteration of the EM algorithm, the proposed LT 
method is henceforth called the iterative Liu-type i.e.,  LT(ITR). 

Unlike the iterative LT method, one can follow \cite{hoerl1975ridge} to estimate the  LT parameters based 
on ridge estimates ${\widehat\sigma}_{R,j}$ and ${\widehat\bb}_{R,j}$. In other words, mixture  
parameter $\bpsi$ and $d_j$ are still iteratively updated in the EM algorithm; however, $k_j,j=1\ldots,J$ parameters 
are estimated only once throughout the EM algorithm using the ridge estimates. Here, we 
estimate  the parameters by ${\widehat k}_{LT,j} = p {\widehat\sigma}_{R,j}/ {\widehat\bb}_{R,j}^\top {\widehat\bb}_{R,j}$. 
This LT estimation method is 
henceforth is called HKP Liu-type i.e., LT(HKP).

 {\bf Liu-type CEM Algorithm}:
 Like previous subsections, the CEM algorithm partitions the observations in the C-step and then update the 
  parameters with in each partition. In the $(r+1)$-th iteration of the CEM algorithm, the E-step remains the same as before. 
 The C-step classifies the observations into partition ${\bf P}^{(r+1)}=(P_1^{(r+1)},\ldots,P_J^{(r+1)})$ where
 $P_j^{(r+1)} = \{({\bf x}_i,y_i); \tau_{ij}(\bpsi^{(r)}) = \underset{h}{\arg\max} ~ \tau_{ih}(\bpsi^{(r)})\}$ with
  $(\tau_{i1}(\bpsi^{(r)},\ldots,\tau_{iJ}(\bpsi^{(r)})$  are obtained from \eqref{tau_ml}.
 Using ${\bf P}^{(r+1)}$, we update the mixing proportions from   \eqref{pihat_cem_ml}.
 We then require to estimate the LT parameters $(k_,d_j)$ in each iteration of the CEM algorithm, 
 Like Liu-type EM algorithm, we propose 
${\widehat k}_{LT,j} = \lambda_{1,j} - 100 \lambda_{p,j}/99$ where $\lambda_{1,j}$ and $\lambda_{p,j}$ 
are the maximum and the minimum 
eigenvalues of ${\bf X}_j^\top {\bf W}_j {\bf X}_j$.

\begin{figure}
\includegraphics[width=1\textwidth]{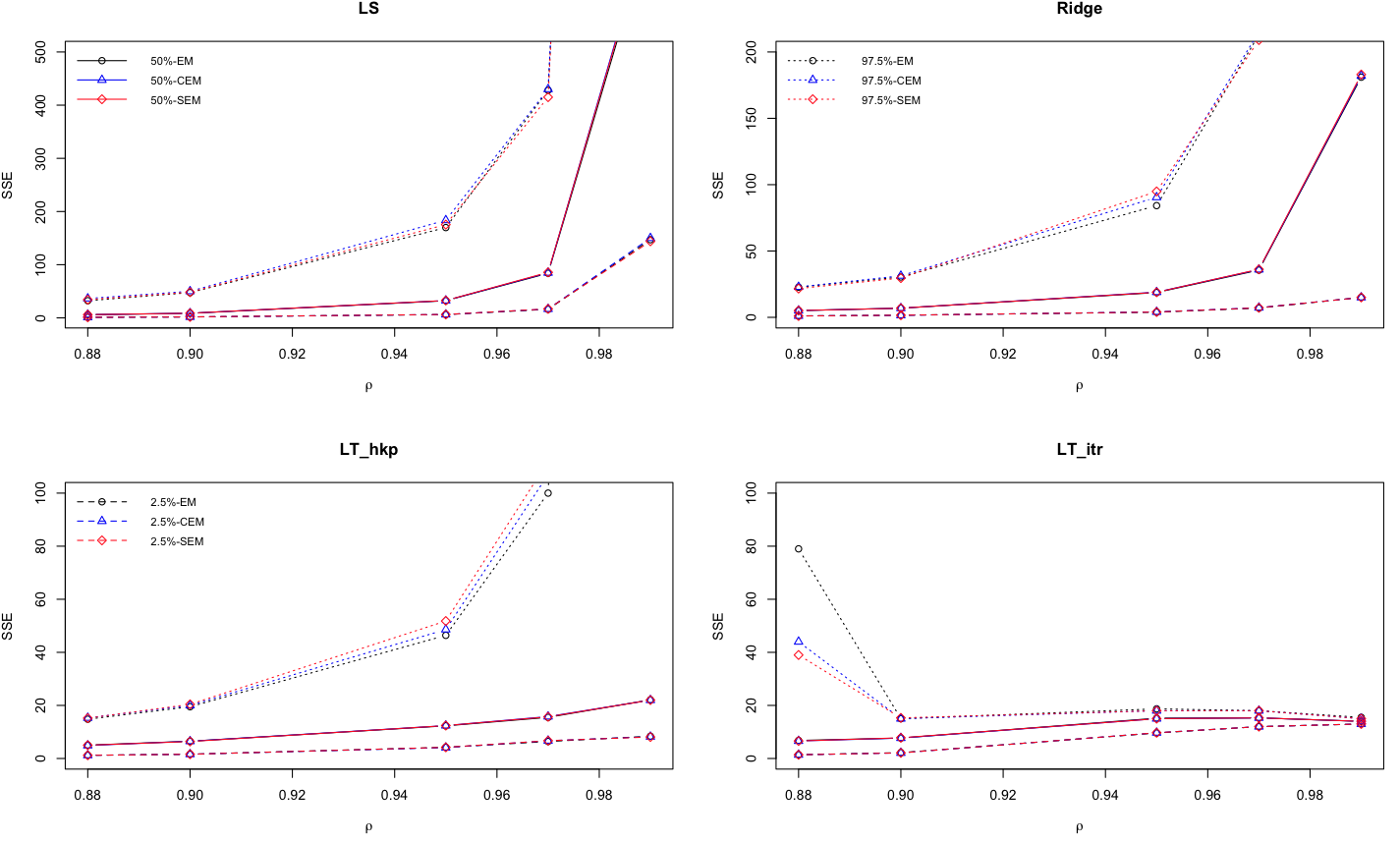}
\caption{The median (M), lower (L) and upper (U) bounds of 95\% CIs for
$\text{SSE}(\widehat{\bf\beta})$ of the estimators when the population is a mixture of two regression models with $n=60$.
}
 \label{beta_J2_n60}
\end{figure}

\begin{lemma} \label{can_cem_lt}
Under the assumptions of Lemma \eqref{can_em_ridge}, the canonical LT estimator in the $j$-th component 
regression $j=1,\ldots,J$ under the CEM algorithm is given by 
 \[
 {\widehat\ba}_{LT,j} = \left( {\bl}_j + k_j\right)^{-1} ({\bl}_j^{1/2} {\bf V}_1^\top {\bf W}_j^{1/2} 
 {\bf y}- d_j {\widehat\ba}_{j}), 
 \]
 and
  \[
  {\widehat\bb}_{LT,j} = {\bf U}_j {\widehat\ba}_{LT,j},
  \]
 where ${\widehat\ba}_{j}$  is the canonical estimate of $\bb_j$ and ${\bf V}_1=[v_{1j},\ldots,v_{pj}]$ with $v_{1j},\ldots,v_{pj}$ are 
 the orthonormal eigenvectors of ${\bf W}_j^{1/2} {\bf X}_j {\bf X}_j^\top {\bf W}_j^{1/2}$.    
 \end{lemma}
From Lemma \ref{can_cem_lt} and Lemma \ref{dj_em}, one can estimate parameter $d_j$ based on partition $P_j^{(r+1)}$ from \eqref{djhat_em} where 
 $(\lambda_{1j},\ldots,\lambda_{pj})$ are eigenvalues of ${\bf X}_j^\top {\bf W}_j {\bf X}_j$ 
and ${\widehat\sigma}_{R,j}^2$ from \eqref{sigj_cem_ridge}.
 To estimate the regression parameters, we implement a WLS based on the LT penalty as 
 \begin{align}\label{wls_cem_lt}
{\widehat \bb}_{LT,j}^{(r+1)} = \underset{\bb_j}{\arg \min} ~ ({\bf y}_j-{\bf X}_j \bb)^\top {\bf W}_j ({\bf y}_j-{\bf X}_j \bb) + 
\left[(-\frac{\widehat{d}_j}{{\widehat k}_{LT,j}^{1/2}}) {\widehat \bb}_j - {\widehat k}_{LT,j}^{1/2} \bb_j \right]^\top \left[(-\frac{\widehat{d}_j}{{\widehat k}_{LT,j}^{1/2}}) {\widehat \bb}_j - {\widehat k}_{LT,j}^{1/2} \bb_j \right],
\end{align}
where ${\bf y}_j$ and ${\bf X}_j$ are response vector and design matrix under $P_j^{(r+1)}$ and ${\widehat \bb}_j$ can be any estimate for ${\bb}_j$. 
Also,  
 ${\bf W}_j$ is a weight diagonal matrix with
 entries
$(\tau_{1j}(\bpsi^{(r)}),\ldots,\tau_{n_j,j}(\bpsi^{(r)}); j=1,\ldots, J$. 
One can easily find the solution to  \eqref{wls_em_lt} and update the regression parameters by
 \begin{align}\label{bej_cem_lt}
{\widehat \bb}_{LT,j}^{(r+1)} = \left({\bf X}_j^\top {\bf W}_j {\bf X}_j + {\widehat k}_{LT,j} \I \right)^{-1} 
({\bf X}_j^\top {\bf W}_j {\bf y}_j-\widehat{d}_j {\widehat \bb}_j),
\end{align}
\begin{align}\label{sigj_cem_lm}
{\widehat\sigma}_{LT,j}^{2(r+1)} = \frac{({\bf y}_j-{\bf X}_j \widehat{\bb}_{LT}^{(r+1)})^\top {\bf W}^{(r)}_j 
({\bf y}_j-{\bf X}_j \widehat{\bb}_{LT}^{(r+1)}) }{\sum_{i=1}^{n} \tau_{ij}(\bpsi^{(r)})}, 
\end{align}
with $\widehat{\bb}_{LT}^{(r+1)}=(\widehat{\bb}_{LT,1}^{(r+1)},\ldots,\widehat{\bb}_{LT,J}^{(r+1)})$.
The E-, C- and M-steps are repeatedly computed until the convergence criterion is satisfied.

\begin{figure}
\includegraphics[width=1\textwidth]{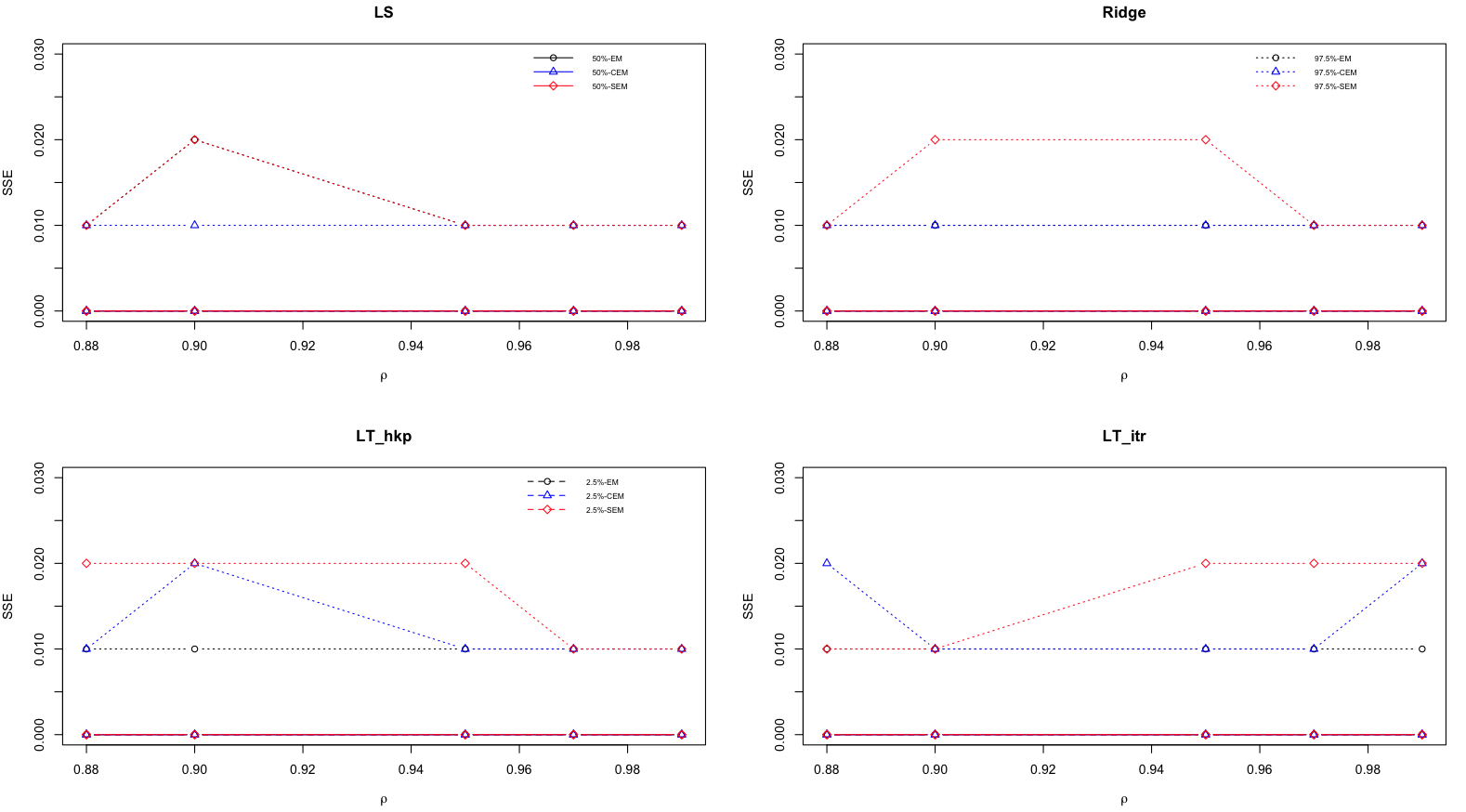}
\caption{The median (M), lower (L) and upper (U) bounds of 95\% CIs for
$\text{SSE}(\widehat{\pi})$ of the estimators when the population is a mixture of two regression models with $n=60$.
}
 \label{pi_J2_n60}
\end{figure}

Unlike the iterative Liu-type CEM algorithm,  one may estimate the parameters of the mixture model via the HKP Liu-type CEM algorithm where the LT  parameters $(k_j,d_j), j=1,\ldots,J$ are updated once throughout the algorithm.
From  \cite{hoerl1975ridge,liu2003using}, we propose to estimate 
${\widehat k}_{LT,j} = p {\widehat\sigma}_{R,j}/ {\widehat\bb}_{R,j}^\top {\widehat\bb}_{R,j}$ 
and ${\widehat d}_j$ from \eqref{djhat_em} where ${\widehat\bb}_{R,j}$ and ${\widehat\sigma}_{R,j}$ come from \eqref{bej_cem_ridge} and \eqref{sigj_cem_ridge}, respectively. 

 {\bf Liu-type SEM Algorithm}:
Similar to the SEM algorithms, the S-step partition the observations stochastically from
$\text{Multi}(1, \tau_{i1}(\bpsi^{(r)}),\ldots,\tau_{iJ}(\bpsi^{(r)}))$ for $i=1,\ldots,n$. 
Once the partition established, the rest of the Liu-type SEM algorithms are implemented similar to the Liu-type CEM algorithms.

\section{Simulation Studies}\label{sec:sim}
In this section, we examine the performance of the ML, Ridge and Liu-type (LT) methods when 
the underlying population is a mixture of linear regression models with multicollinearity 
problem. We present two simulations enabling us to study the effect of the sample size, 
multicollinearity levels and the number of components of mixture models on the estimation 
and prediction of the proposed methods. 

\begin{figure}
\includegraphics[width=1\textwidth]{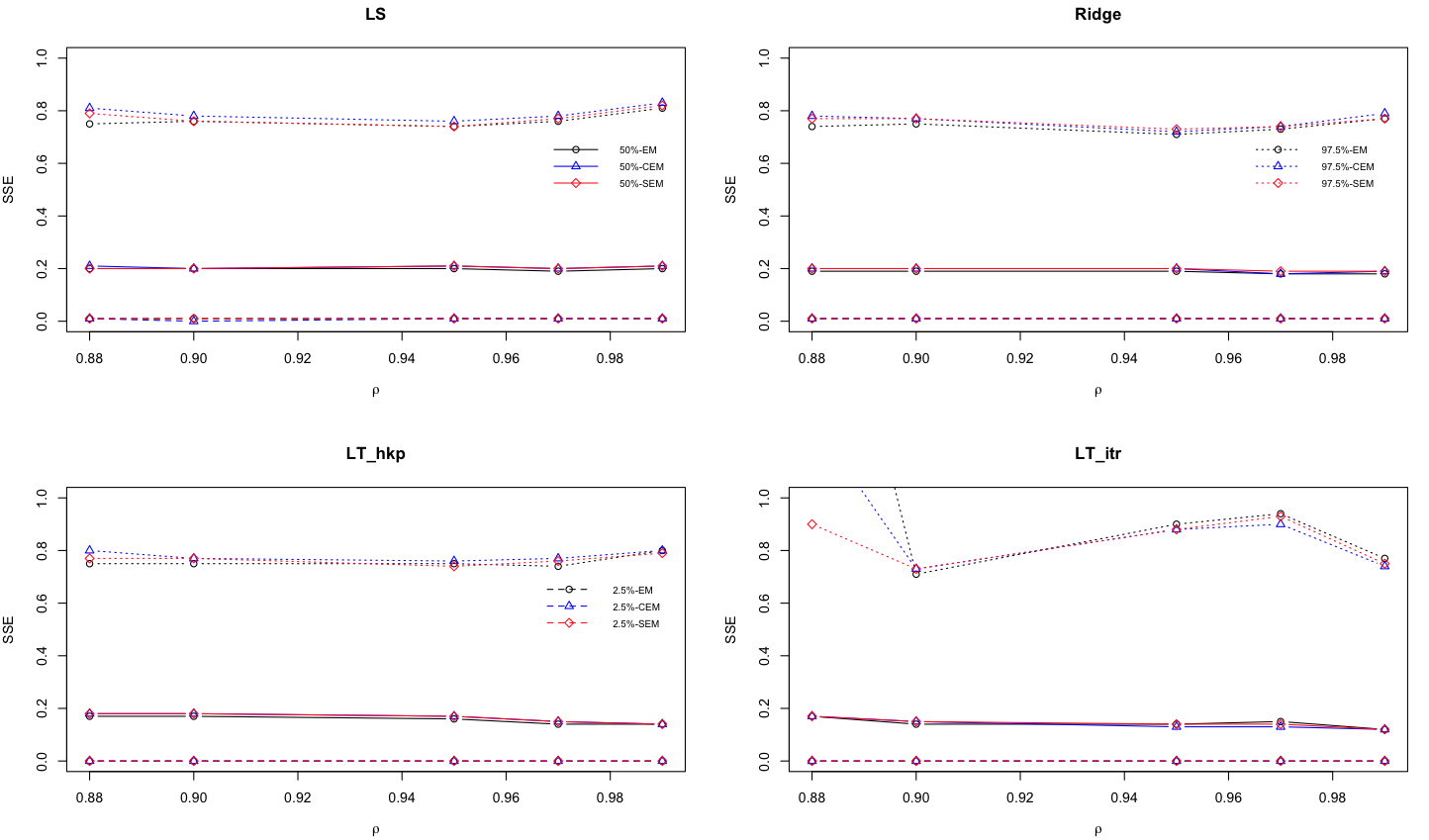}
\caption{The median (M), lower (L) and upper (U) bounds of 95\% CIs for
$\text{SSE}(\widehat{\sigma}^2)$ of the estimators when the population is a mixture of two regression models with $n=60$.
}
 \label{sig_J2_n60}
\end{figure}

\begin{table}
\label{Pre_2Mix_size60}
\caption{The median (M) and the length (L) of 95\% CIs for the RMSEP of the ML, ridge and LT methods 
in predicting the mixture of two regression models when $n=60$.}
\begin{centering}
\begin{tabular}{ccccccccccccccccc}
\cline{1-17} 
 & $\rho$ & \multicolumn{1}{c}{} & \multicolumn{2}{c}{0.88} &  & \multicolumn{2}{c}{0.90} &  & \multicolumn{2}{c}{0.95} & 
  & \multicolumn{2}{c}{0.97} &  & \multicolumn{2}{c}{0.99}\tabularnewline
\cline{4-5} \cline{7-8} \cline{10-11} \cline{13-14} \cline{16-17} 
Method & Algorithm &  & M & L &  & M  & L &  & M & L &  & M & L &  & M & L\tabularnewline
\hline 
ML & EM &  & 16.3 & 10.8 &  & 16.7 & 11.1 &  & 17.7 & 12.5 &  & 18.1 & 12.1 &  & 18.3 & 12.3\\[-1.5ex]
 & CEM &  & 16.5 & 10.7 &  & 16.7 & 11.1 &  & 17.7 & 12.0 &  & 18.0 & 12.0 &  & 18.2 & 12.3\\[-1.5ex]
 & SEM &  & 16.3 & 11.3 &  & 16.7 & 11.4 &  & 17.6 & 11.9 &  & 17.9 & 11.9 &  & 18.2 & 12.4\\
\hline 
Ridge & EM &  & 16.5 & 11.0 &  & 16.7 & 11.2 &  & 17.9 & 12.1 &  & 18.1 & 12.0 &  & 18.3 & 12.0\\[-1.5ex]
 & CEM &  & 16.4 & 10.8 &  & 16.7 & 11.1 &  & 17.8 & 11.6 &  & 17.9 & 11.9 &  & 18.3 & 12.7\\[-1.5ex]
 & SEM &  & 16.4 & 11.0 &  & 16.7 & 11.5 &  & 17.7 & 11.6 &  & 18.0 & 11.8 &  & 18.3 & 12.3\\
\hline 
LT(HKP) & EM &  & 16.4 & 10.6 &  & 16.6 & 11.1 &  & 17.7 & 11.7 &  & 18.1 & 12.5 &  & 18.3 & 12.0\\[-1.5ex]
 & CEM &  & 16.4 & 11.3 &  & 16.6 & 11.2 &  & 17.6 & 11.9 &  & 18.1 & 11.6 &  & 18.3 & 12.2\\[-1.5ex]
 & SEM &  & 16.4 & 10.9 &  & 16.6 & 11.2 &  & 17.8 & 12.0 &  & 18.0 & 12.1 &  & 18.3 & 12.1\\
\hline 
LT(ITE) & EM &  & 16.4 & 10.8 &  & 16.8 & 10.9 &  & 17.3 & 11.4 &  & 17.7 & 11.8 &  & 18.1 & 12.0\\[-1.5ex]
 & CEM &  & 16.5 & 10.9 &  & 16.7 & 10.9 &  & 17.2 & 11.7 &  & 17.7 & 11.8 &  & 18.1 & 11.9\\[-1.5ex]
 & ESM &  & 16.4 & 11.1 &  & 16.7 & 11.0 &  & 17.2 & 11.7 &  & 17.7 & 12.3 &  & 18.1 & 11.9\\
\hline 
\end{tabular}
\par\end{centering}
\end{table}

In the first study, we simulate data from a population corresponding to a mixture of two 
regression models with four covariates $(x_1,\ldots,x_4)$.  
Following  \cite{inan2013liu}, we use $\rho$ denoting the correlation between covariates to 
simulate the multicollinearity in the mixture model. 
To do so, we first generate random numbers $\{w_{ij};i=1,\ldots,n;j=1,\ldots,5\}$ from the standard normal distribution.  
The covariates are then generated by
\[
x_{ij} = (1-\rho^2)^{1/2} w_{ij} + \rho w_{i,5}, ~~~ j=1,\ldots,4,
\]
where we set $\rho=\{0.88,0.9,0.95,0.97,0.99\}$ to simulate the multicollinearity levels in the
 mixture of regression models. 
 The responses are then generated from mixture model \eqref{mix_reg} whose true parameters are given 
 by ${\bf \Psi}_0 = (\pi_0,{\bf \beta}_{01},{\bf \beta}_{02},\sigma^2_{01},\sigma^2_{02})$ with
  $\pi_0=0.7$, ${\bf \beta}_{01}=(1,3,4,5,6)$
  , ${\bf \beta}_{02}=(-1,-1,-2,-3,-5)$ and $\sigma^2_{01}=\sigma^2_{02}=1$.
  We measured the estimation performance of the ML, Ridge and LT methods by sum of squared errors (SSE) in estimating 
  $(\pi,{\bf \beta},\sigma^2)$ and computed 
  $\text{SSE}(\widehat{\bf\beta}) = \left[ (\widehat{\bf\beta} - {\bf\beta}_0)^\top (\widehat{\bf\beta} - {\bf\beta}_0)\right]$,
  $\text{SSE}({\widehat\pi}) = (\widehat{\pi} -\pi_0)^2$ and 
  $\text{SSE}(\widehat{\sigma}^2) = \left[ (\widehat{\sigma}^2 - {\sigma_0^2})^\top (\widehat{\sigma}^2 - {\sigma_0^2})\right]$,
 where $\widehat{\bf\beta}=(\widehat{\bf\beta}_1 , \widehat{\bf\beta}_2)$,
  $\widehat{\sigma}^2=({\widehat\sigma}_1^2, {\widehat\sigma}_2^2)$,
  $\widehat{\bf\beta}_0=(\widehat{\bf\beta}_{01}, \widehat{\bf\beta}_{02})$
  and ${\sigma_0^2}=({\sigma_{01}^2}, {\sigma_{02}^2})$.
  We also used the root mean squared errors of prediction (RMSEP) to evaluate the prediction performance of the methods. 
  To do so, we first compute the RMSEP of the $m$-th replicate through $K$-fold cross-validation by 
  \[
  \text{RMSEP}^{(m)} = \left(1/n \sum_{i=1}^{n} (y_i -\widehat{y}_i^{(m)})^2 \right)^{1/2},
  \]
  where $\widehat{y}_i^{(m)}$ is the predicted response of the $i$-th observation in the $m$-th replicate.
  
  We computed the estimation and prediction measures for the ML, Ridge and LT methods as follows. 
  We first generated a sample of size $n=\{60,100\}$ from the underlying mixture of regression 
  models as described above. We then used the EM, CEM and SEM algorithms to estimate the parameters 
  of the mixture population via ML, ridge and LT methods. We applied the idea of $K=5$ cross-validation 
  to assess the prediction performance of the methods. To this end, we divided the sample into $K$ 
  folds of equal sizes. We used $K-1$ folds for training and the remaining fold for prediction. 
  We repeated the procedure for all $k=1,\ldots,K$ to compute the $\text{RMSEP}^{(m)}$. Eventually, 
  we replicated the entire procedures $m=2000$ times and computed the median and 95\% confidence 
  interval (CI) for the SSE and RMSEP measures. The lower and upper bounds of the CI correspond 
  to 2.5 and 97.5 percentiles of 2000 replications, respectively. 
  
  \begin{figure}
\includegraphics[width=1\textwidth]{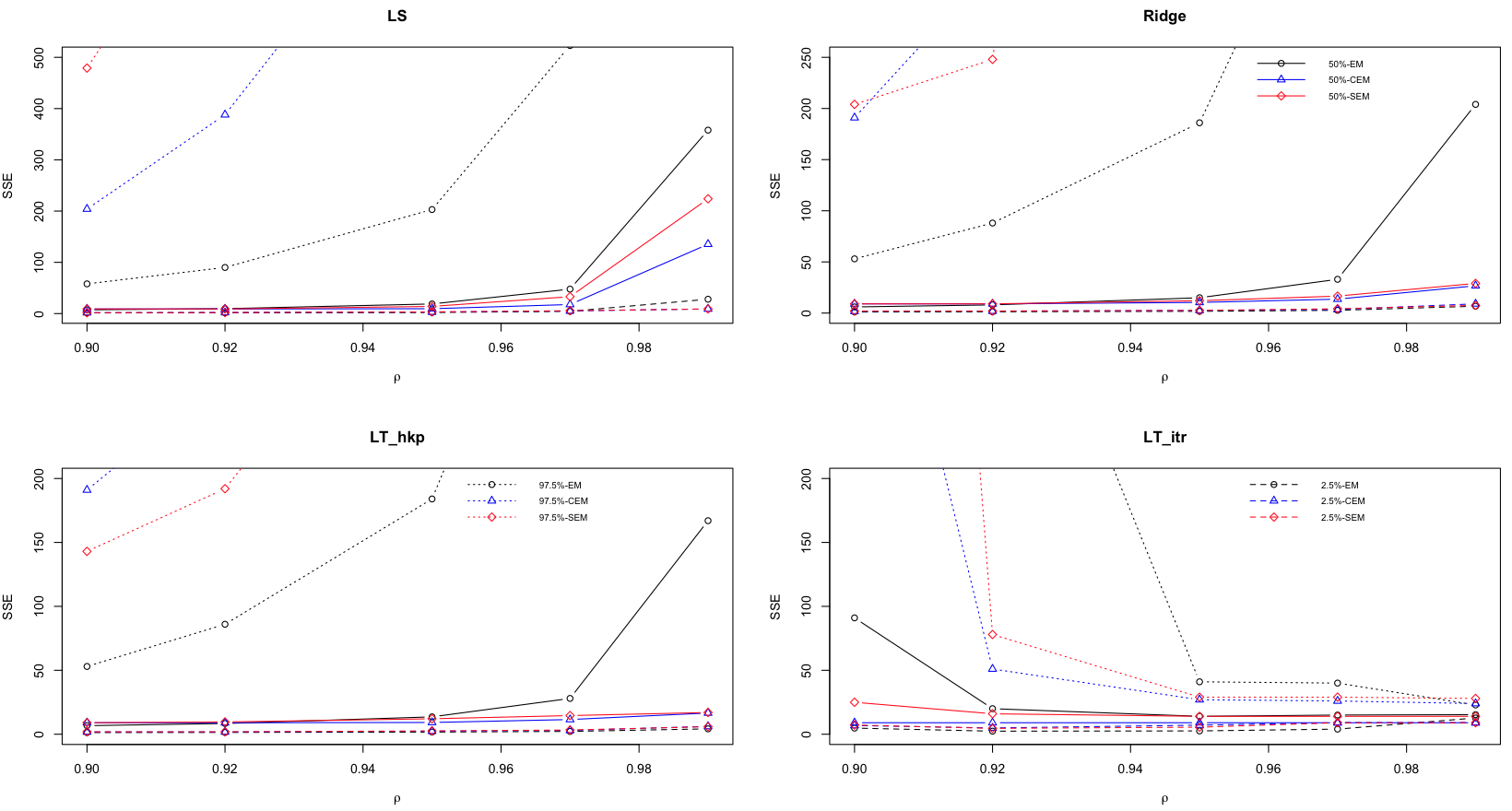}
\caption{The median (M), lower (L) and upper (U) bounds of 95\% CIs for
$\text{SSE}(\widehat{\pi})$ of the estimators when the population is a mixture of three regression models with $n=60$.
} \label{beta_J3_n60}
\end{figure}

\begin{figure}
\includegraphics[width=1\textwidth]{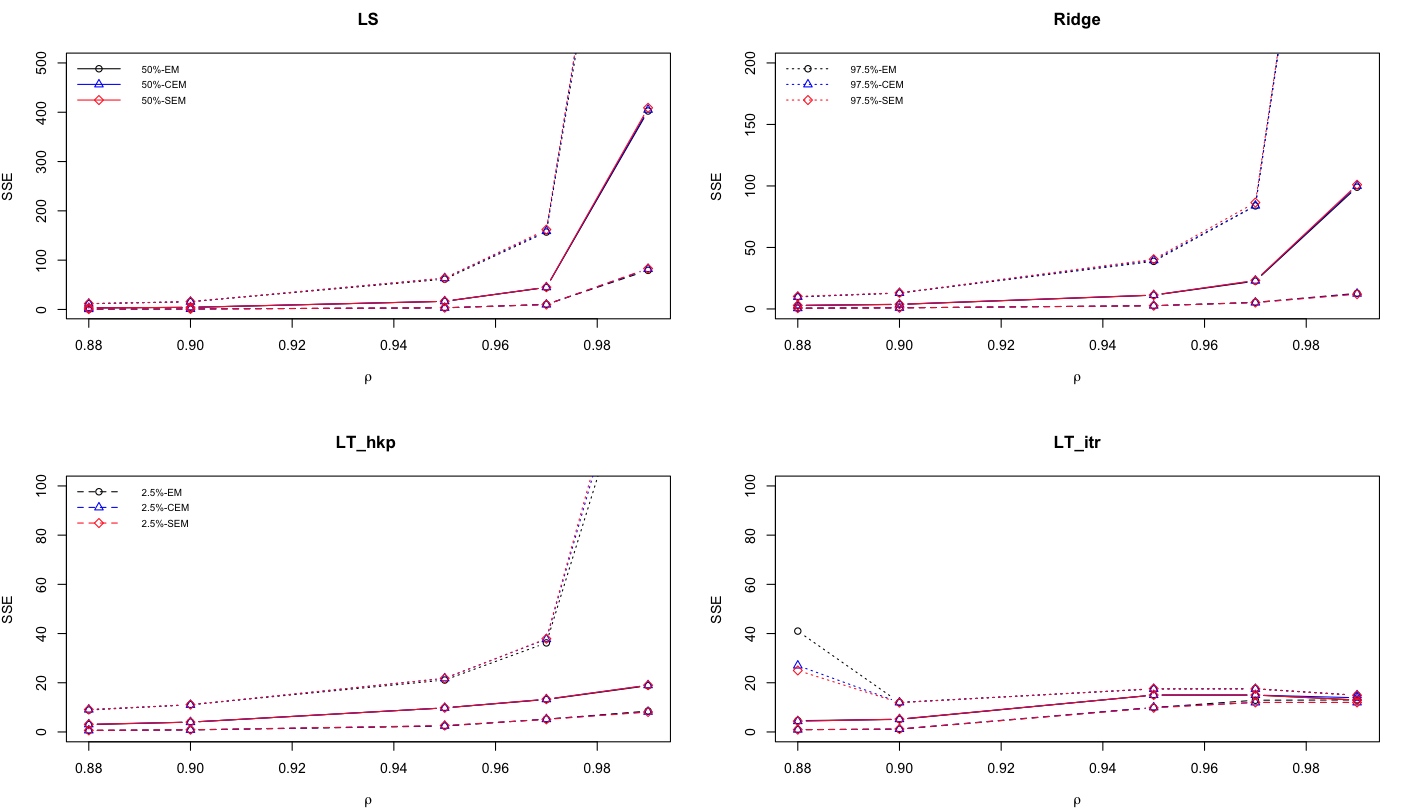}
\caption{The median (M), lower (L) and upper (U) bounds of $95\%$ CIs for
$\text{SSE}(\widehat{\bf\beta})$ of the estimators when the population is a mixture of two regression models with $n=100$.
}
 \label{beta_J2_n100}
\end{figure}

\begin{figure}
\includegraphics[width=1\textwidth]{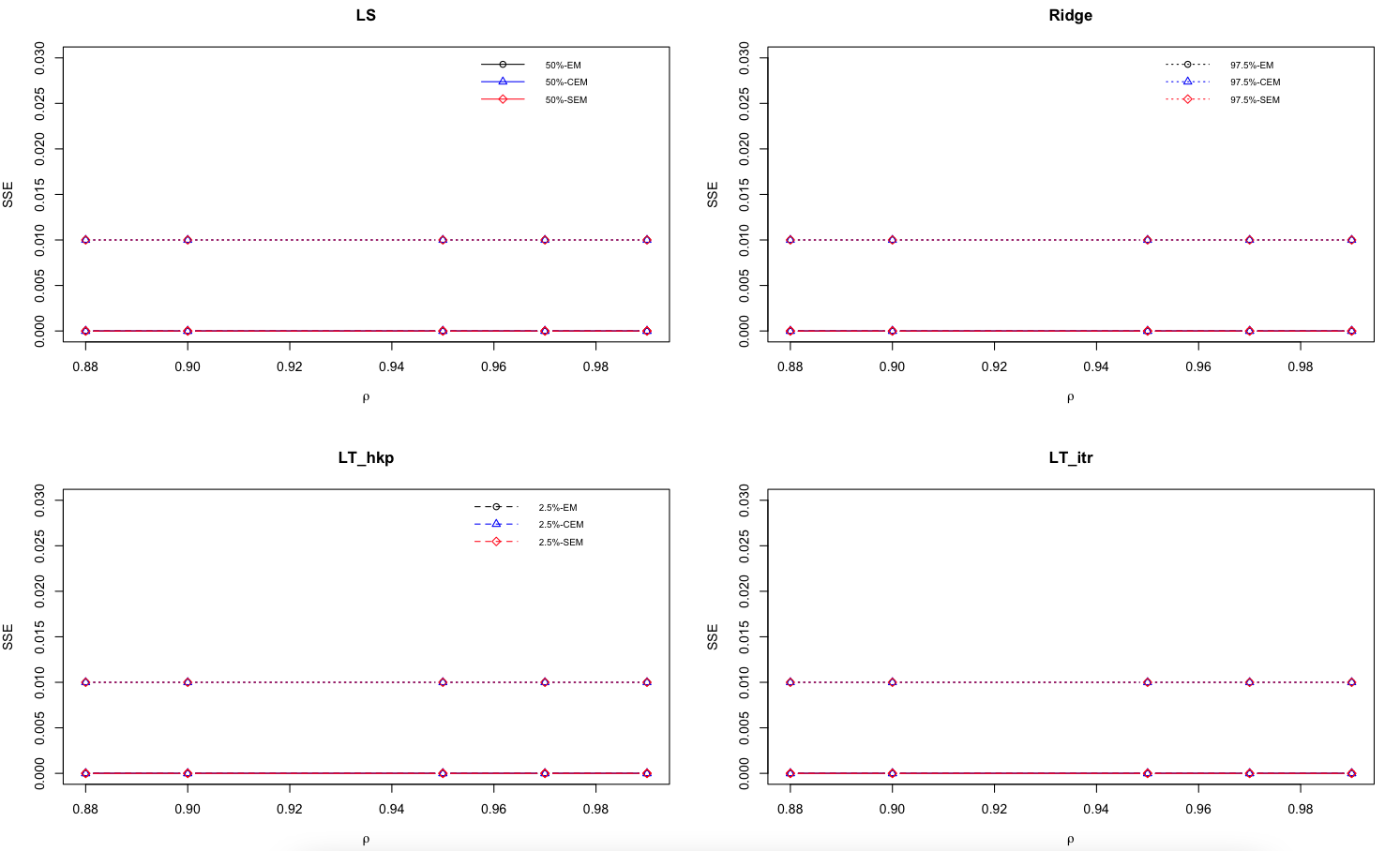}
\caption{The median (M), lower (L) and upper (U) bounds of $95\%$ CIs for
$\text{SSE}(\widehat{\pi})$ of the estimators when the population is a mixture of two regression models with $n=100$.
} \label{pi_J2_n100}
\end{figure}

\begin{figure}
\includegraphics[width=1\textwidth]{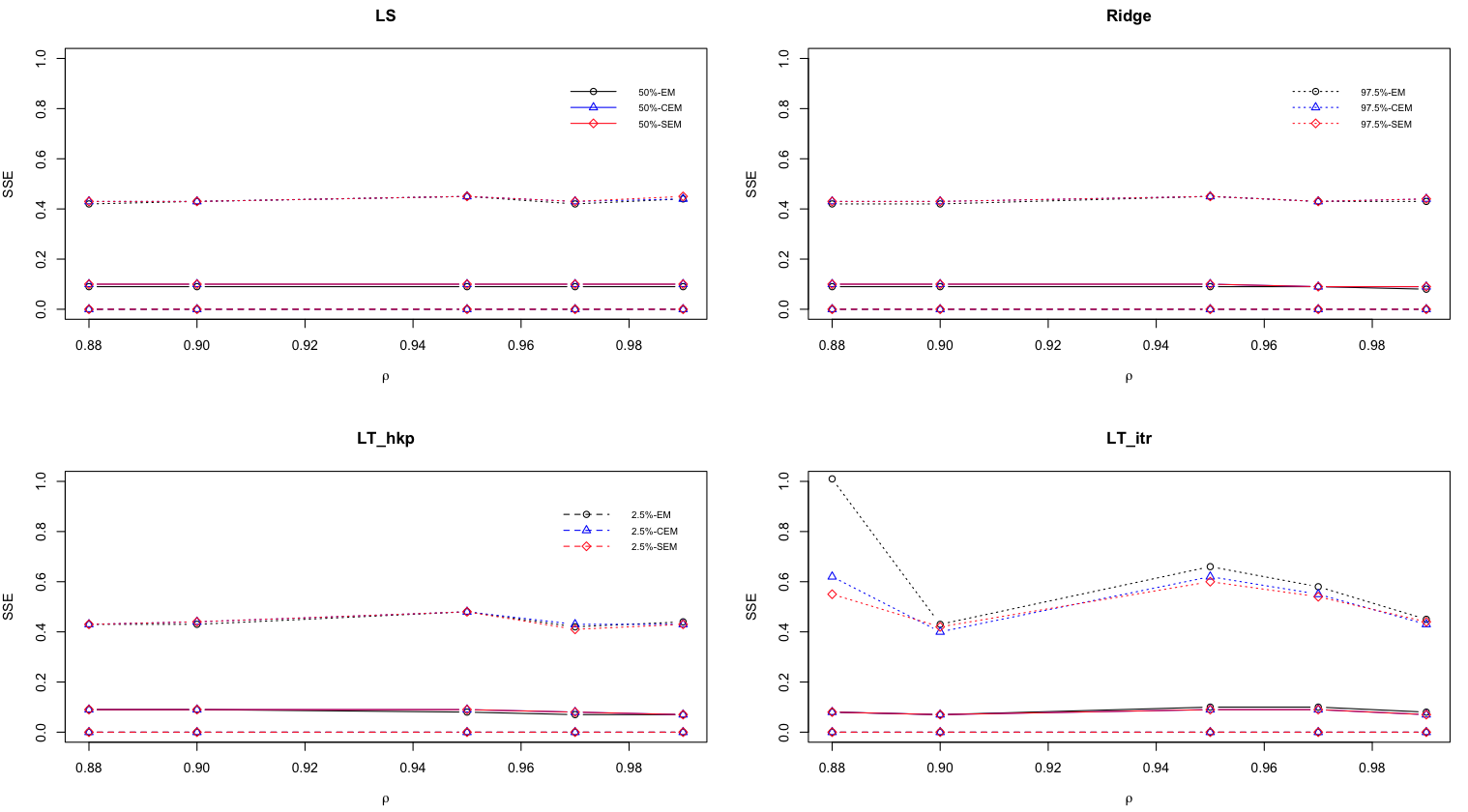}
\caption{The median (M), lower (L) and upper (U) bounds of $95\%$ CIs for
$\text{SSE}({\widehat\sigma}^2)$ of the estimators when the population is a mixture of two regression models with $n=100$.
} \label{sig_J2_n100}
\end{figure}

  We show the results of the simulation study in estimating $(\pi,{\bf \beta},{\bf \sigma}^2)$ 
  in Figures \ref{beta_J2_n60}-\ref{sig_J2_n100}. It is observed that the ML methods estimate 
  slightly better the mixing proportion than the ridge and LT methods. This happens because the 
  ridge and LT estimators are biased shrinkage methods where a slight bias is incorporated into 
  the estimation to encounter the multicollinearity problem. We observe 
  that the multicollinearity significantly impacts the ML estimates of the coefficients and results 
  in extremely unreliable estimates for all EM, CEM and SEM algorithms. Unlike ML estimates, 
  a significant improvement is seen in the performance of the shrinkage methods in estimating the 
  coefficients of the component regressions. The LT methods appear more reliable than 
  their ridge counterparts in the multicollinearity. From a comparison between 
  LT(ITR)  and LT(HKP), we see that LT(HKP) provides more reliable estimates for ${\bf\sigma}^2$. 
  Among the LT(HKP) estimators, the CEM algorithm almost always outperforms its EM ad SEM counterparts. 
  Tables \ref{Pre_2Mix_size60}-\ref{Pre_2Mix_size100} show the median and the length of 95\% CI of the  
  RMSEP for all the developed methods. The tables clearly show that the prediction performances of all 
  the methods and EM algorithms  are almost identical. This finding is consistent with 
  \cite{inan2013liu,elsayed_log} that multicollinearity seriously affects the estimation of the methods 
  while prediction levels stay almost the same.  
   
   \begin{table}
\caption{\footnotesize{The median (M), lower (L) and upper (U) bounds of 95\% CIs for $\sqrt{\text{SSE}}$ of the methods 
in the analysis of bone mineral data with sample size $n=60$.}}
\label{bone_1_size60}
\begin{centering}
\begin{tabular}{cccccccccccccc}
\hline 
 &  &  & \multicolumn{3}{c}{CEM} &  & \multicolumn{3}{c}{SEM} &  & \multicolumn{3}{c}{EM}\tabularnewline
 \cline{4-6} \cline{8-10} \cline{12-14}
Methods & $\ensuremath{\ensuremath{\boldsymbol{\Psi}}}$ &  & M & L & U &  & M  & L & U &  & M & L & U\tabularnewline
\hline 
ML & $\ensuremath{\beta}$ &  & .010 & .002 & .165 &  & .019 & .003 & .213 &  & .018 & .003 & .134\\ [-1.5ex]
 & $\ensuremath{\pi}$ &  & .333 & .100 & .366 &  & .333 & .183 & .366 &  & .218 & .015 & .365\\ [-1.5ex]
 & $\sigma^{2}$ &  & .003 & .000 & .014 &  & .006 & ..000 & .014 &  & .004 & .000 & .014\\ 
\hline 
Ridge & $\ensuremath{\beta}$ &  & .009 & .002 & .165 &  & .013 & .002 & .166 &  & .012 & .002 & .118\\ [-1.5ex] 
 & $\ensuremath{\pi}$ &  & .333 & .100 & .366 &  & .333 & .166 & .366 &  & .214 & .019 & .366\\ [-1.5ex]
 & $\sigma^{2}$ &  & .003 & .000 & .014 &  & .006 & .000 & .014 &  & .004 & .000 & .014\\ 
\hline 
LT(HKP) & $\ensuremath{\beta}$ &  & .009 & .002 & .165 &  & .010 & .003 & .183 &  & .010 & .003 & .067\\ [-1.5ex]
 & $\ensuremath{\pi}$ &  & .333 & .100 & .366 &  & .333 & .150 & .366 &  & .205 & .013 & .372\\ [-1.5ex]
 & $\sigma^{2}$ &  & .003 & .000 & .014 &  & .006 & .000 & .014 &  & .004 & .000 & .016\\ 
\hline 
LT(ITE) & $\ensuremath{\beta}$ &  & .009 & .002 & .010 &  & .009 & .006 & .011 &  & .009 & .007 & .010\\ [-1.5ex]
 & $\ensuremath{\pi}$ &  & .300 & .100 & .366 &  & .350 & .116 & .566 &  & .575 & .032 & .599\\ [-1.5ex]
 & $\sigma^{2}$ &  & .002 & .000 & .014 &  & .005 & .000 & .014 &  & .003 & .000 & .009\\ 
\hline 
\end{tabular}
\par\end{centering}
\end{table}

\begin{figure}
\includegraphics[width=1\textwidth]{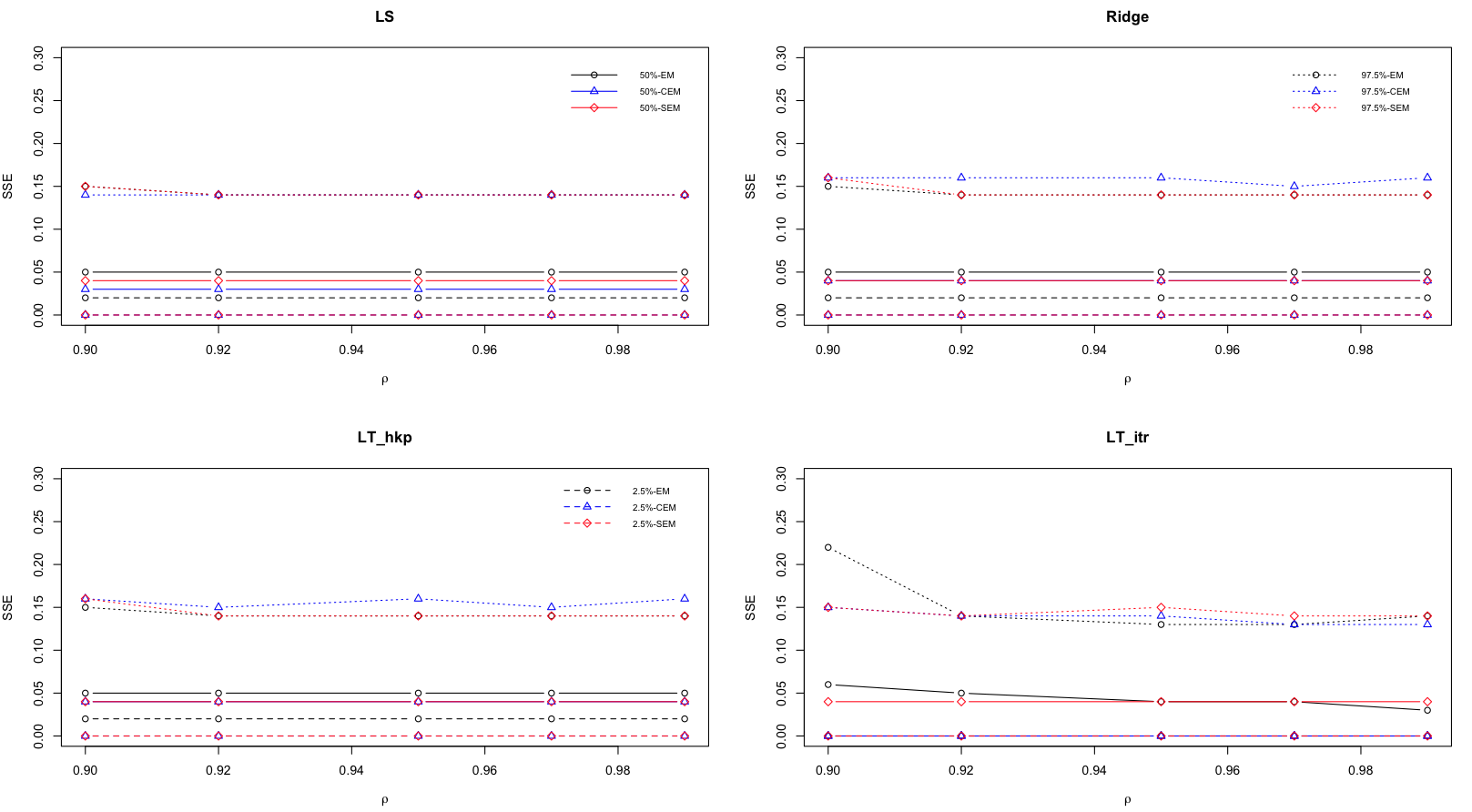}
\caption{The median (M), lower (L) and upper (U) bounds of $95\%$ CIs for
$\text{SSE}(\widehat{\pi})$ of the estimators when the population is a mixture of three regression models with $n=100$.
} \label{pi_J3_n60}
\end{figure}

\begin{figure}
\includegraphics[width=1\textwidth]{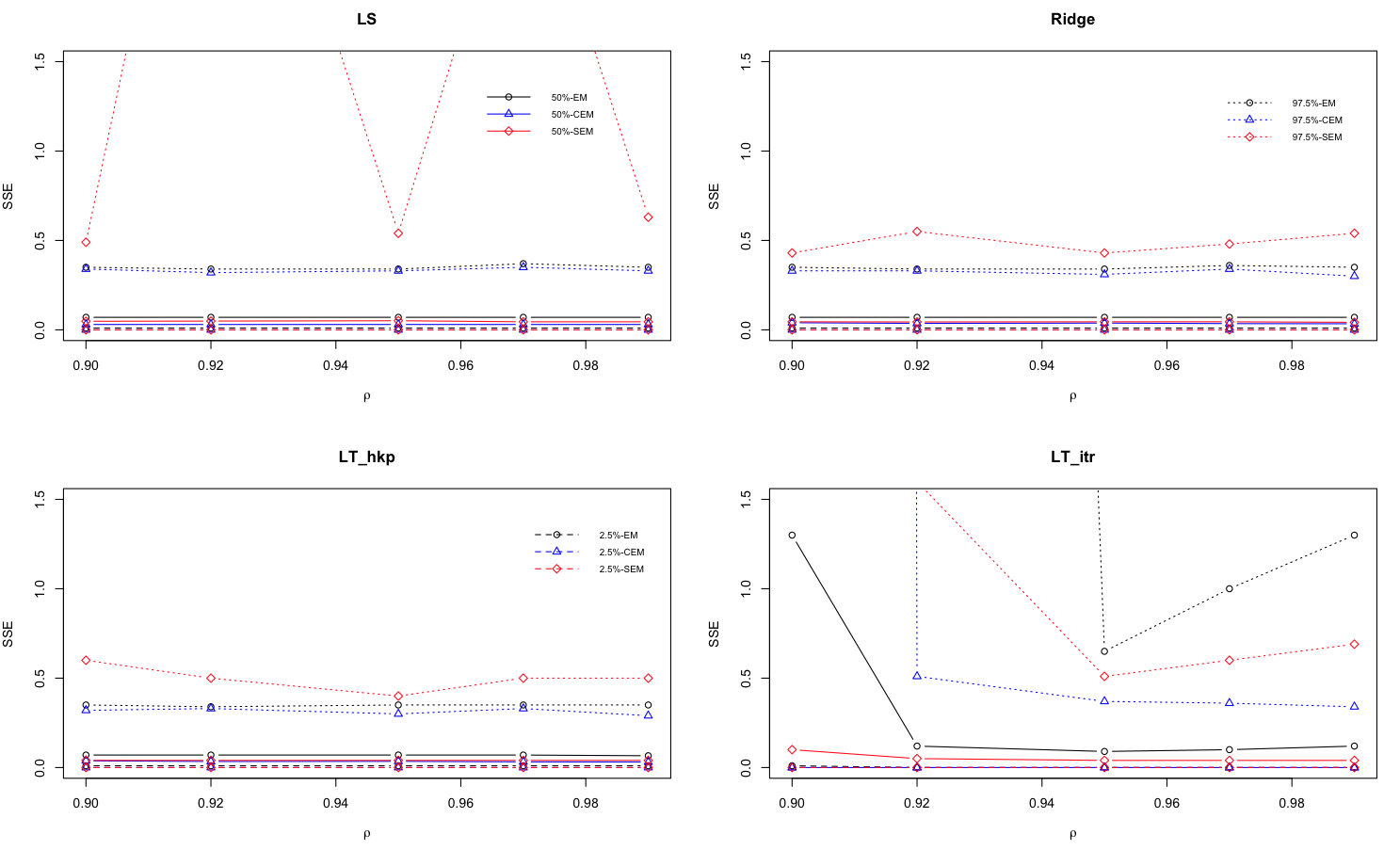}
\caption{The median (M), lower (L) and upper (U) bounds of 95\% CIs for
$\text{SSE}({\widehat\sigma}^2)$ of the estimators when the population is a mixture of three regression models with $n=100$.
} 
 \label{sig_J3_n60}
\end{figure}

\begin{figure}
\includegraphics[width=1\textwidth]{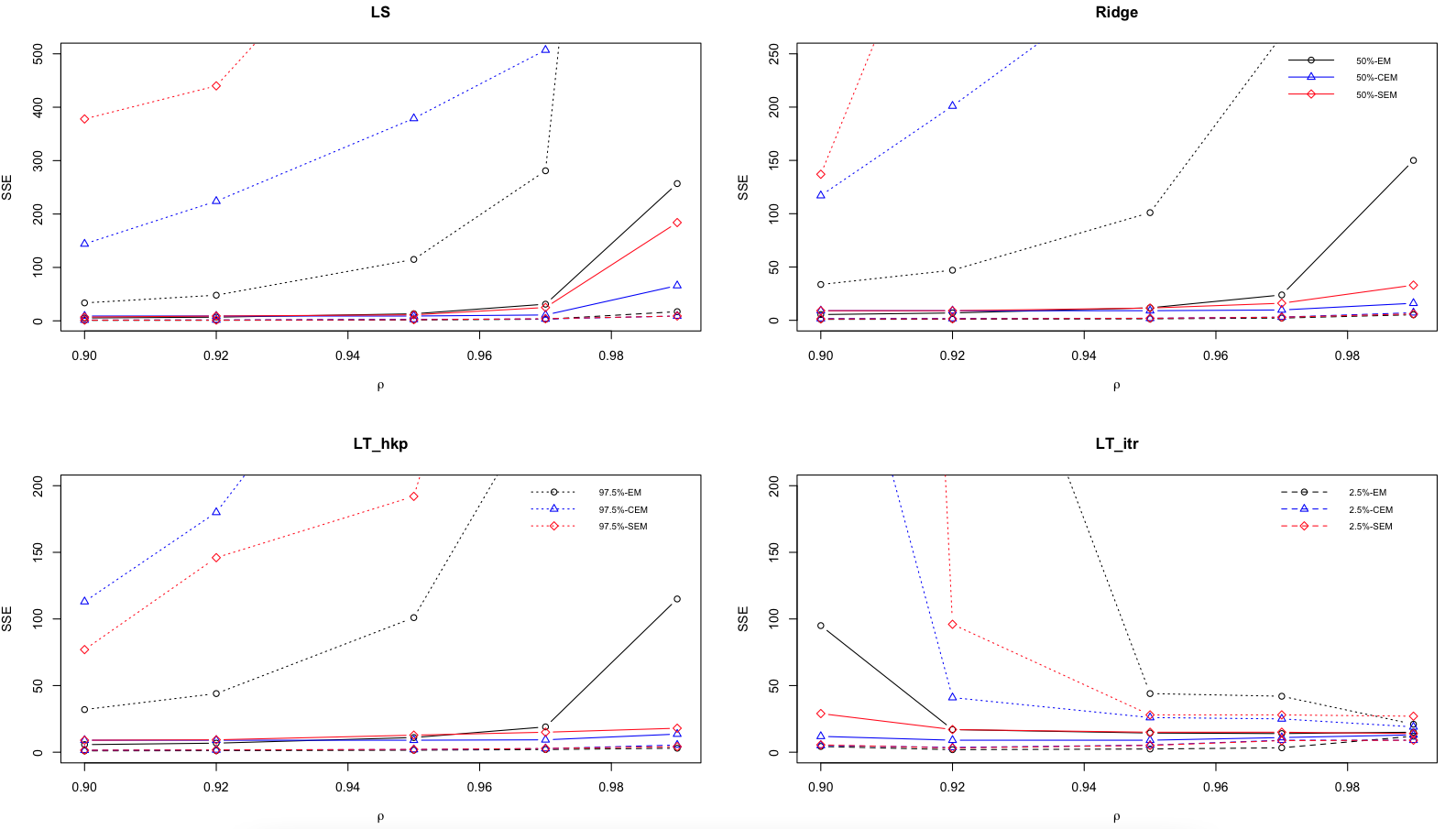}
\caption{The median (M), lower (L) and upper (U) bounds of 95\% CIs for
$\text{SSE}(\widehat{\bf\beta})$ of the estimators when the population is a mixture of three regression models with $n=100$.
}  
\label{beta_J3_n100}
\end{figure}

\begin{figure}
\includegraphics[width=1\textwidth]{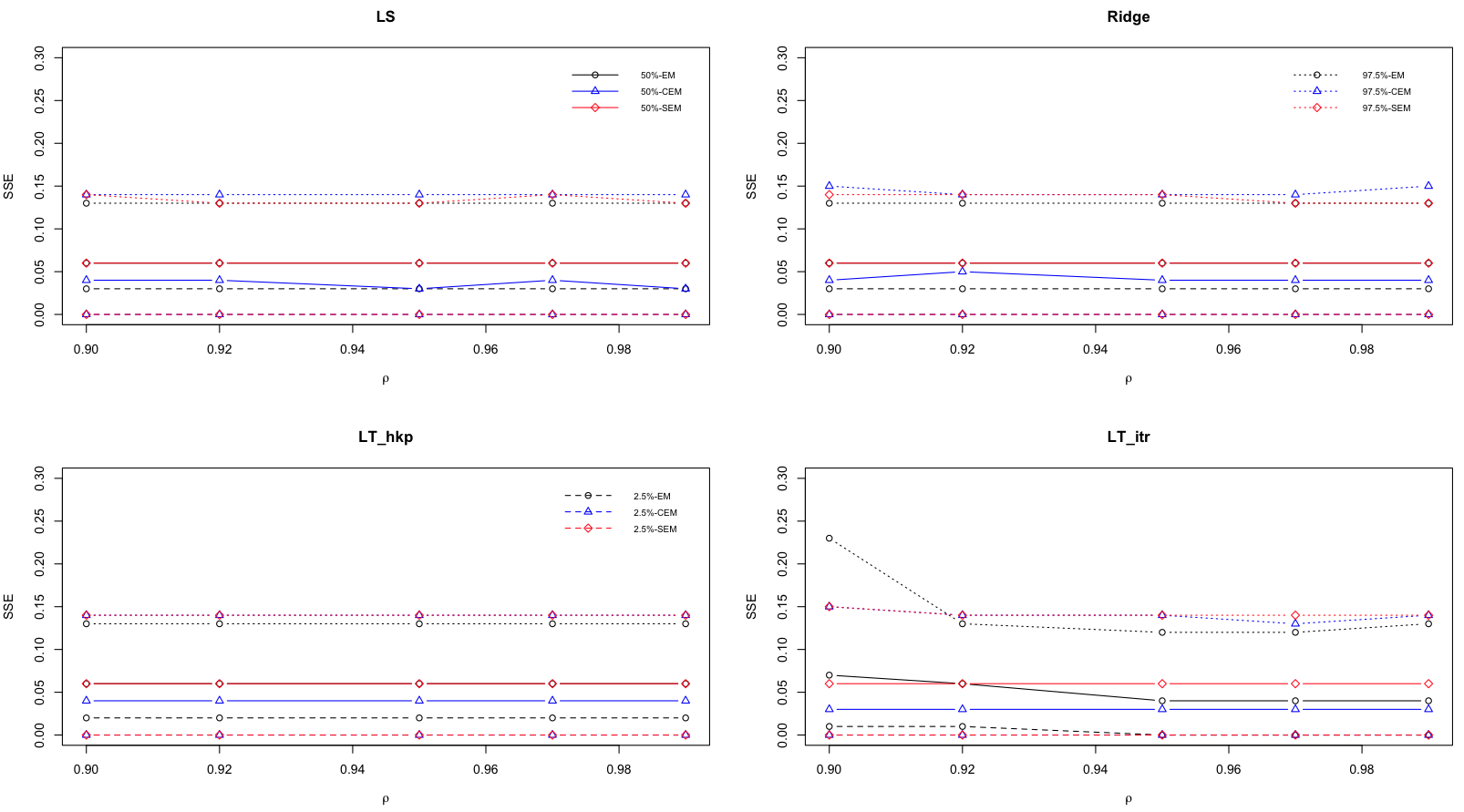}
\caption{The median (M), lower (L) and upper (U) bounds of 95\% CIs for
$\text{SSE}(\widehat{\pi})$ of the estimators when the population is a mixture of three regression models with $n=100$.
}   \label{pi_J3_n100}
\end{figure}

\begin{figure}
\includegraphics[width=1\textwidth]{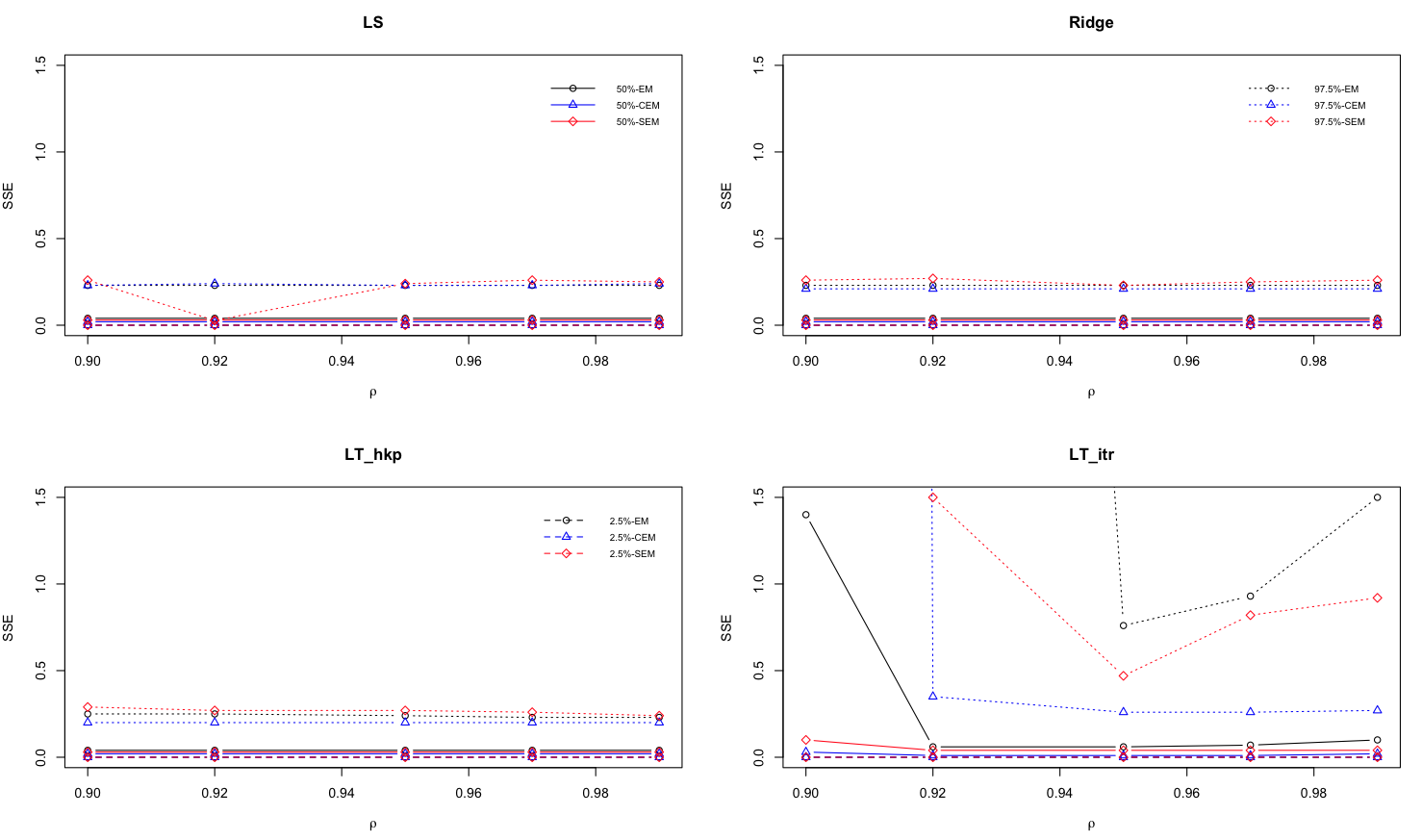}
\caption{The median (M), lower (L) and upper (U) bounds of 95\% CIs for
$\text{SSE}({\widehat\sigma}^2)$ of the estimators when the population is a mixture of three regression models with $n=100$.
}   \label{sig_J3_n100}
\end{figure}

  In the second simulation study, we evaluate the performance of the estimators when the 
  underlying mixture model consists of three component regressions with two covariates. 
  We set $\rho=\{0.9,0.92,0.95,0.97,0.99\}$ to simulate the multicollinearity in the mixture. 
  As described earlier, we generated the covariates and responses from the mixture model with 
  parameters $\pi_0=(\pi_{01},\pi_{02},\pi_{03})=(0.3,0.4,0.3)$, ${\bf \beta}_{01}=(1,3,4)$, 
  ${\bf \beta}_{02}=(-1,-1,-2)$, ${\bf \beta}_{03}=(-3,1,-4)$ and $\sigma^2_{0}=(0.25,1,0.09)$. 
  Similar to the settings of the first simulation study, we replicated 2000 times all the estimation 
  and prediction procedures under the EM, CEM and SEM algorithms and computed the median and 95\% CI 
  for the SSEs and RMSEP for size sizes $n=\{60,100\}$. The estimation and prediction results are 
  demonstrated in Figures \ref{beta_J3_n60}-\ref{sig_J3_n100} and Tables \ref{Pre_3Mix_size60}-\ref{Pre_3Mix_size100}, respectively. 
  Here, we also observe that the ML method slightly better estimates the mixing proportions; however, 
  the ML methods result in extremely unreliable estimates for the coefficients of component regressions. 
  It is easy to see that shrinkage estimators do better in estimating the component regression parameters. 
  In addition, the LT(HKP) almost always outperforms other methods and provides a more reliable estimate 
  of the mixture of regression models. Therefore, the LT(HKP)  method based on the CEM algorithm is 
recommended to fit the mixture of linear regression models in multicollinearity.

\section{Bone Data Analysis}\label{sec:real}
As a bone metabolic disease, osteoporosis occurs when the bone mineral architecture of the body 
deteriorates. This deterioration results in skeletal fragility and a high risk of osteoporotic 
fractures in different body 
areas such as the hip and femur. Osteoporosis and its related diseases significantly impact the 
patient's health and survival. For example, one out of every two patients with osteoporotic hip 
fractures can no longer live independently, and one out of three may die within one year after 
the medical complication of the broken bone \citep{bliuc2009mortality,neuburger2015impact}. 
The bone mineral density (BMD) of individuals improves until age 30 and then decreases as 
individuals age. The BMD score of an individual is compared with a BMD norm to determine 
the bone disorder status. The BMD norm is computed by the mean BMD scores of healthy adults 
aged 20-30. The bone status of an individual is diagnosed as osteoporosis when the BMD score 
is less than -2.5 SD from the BMD norm of the population.

The BMD scores are obtained vis an expensive and time-consuming procedure. Despite this, the
 researchers have access to various easy-to-measure patients' characteristics, such as age,
  weight, BMI, and test results from earlier surveys \citep{kim2012relationship,felson1993effects}. Regression models 
  are among the most 
  common methods to investigate the impact of a set of patients' characteristics on the  BMD 
  responses. The impact of the characteristics may differ at different BMD levels. Hence, the 
  inference on BMD measurements can be handled as a problem of the mixture of linear regression models.

The bone mineral data in this section were obtained from the National Health and Nutritional 
Examination Survey (NHANES III) administrated by the CDC on more than  33999 American adults 
between years 1988 to 1994. One hundred eighty-two white women aged 50 and older participated 
in all two bone examinations during the survey. Owing to the significant impact of osteoporosis 
on older women, we treated these 182 women as our underlying population. We considered the femur 
BMD  from the second bone examination as the response variable. 
We also used two easily attainable physical characteristics as the explanatory variables of the 
regression. These physical characteristics include arm and bottom circumferences. The association 
between the two explanatory variables is $\rho=0.81$, indicating the multicollinearity in the 
regression. Using the ML method based on the information of all individuals in the population, 
the BIC criterion suggests that a mixture of two regression models is the best fit with parameters 
$\pi_0=(\pi_{01},\pi_{02})=(0.64,0.36)$, ${\bf \beta}_{01}=(0.004,0.006)$, 
  ${\bf \beta}_{02}=(0.011,0.005)$ and $\sigma^2_{0}=(0.0081,0.0144)$.
These parameters were then considered as the true parameters of the mixture population. 
We replicated 2000 times the ML, ridge and LT methods under EM, CEM and SEM algorithms in 
estimating the parameters of the bone mineral population based on sample sizes  $n=\{60,100\}$. 
We then computed the estimation and prediction measures 
$\sqrt{\text{SSE}(\widehat{\bf\beta})},  \sqrt{\text{SSE}(\widehat{\pi})}, \sqrt{\text{SSE}(\widehat{\sigma^2})}$ and 
MRSEP using 5-fold cross-validation as described in Section \ref{sec:sim}.

Tables \ref{bone_1_size60}-\ref{MRSEP_Real} report the median and 95\% CI for the above measures in
 estimating and predicting 
the bone mineral population. The CI's lower (L) and upper (U) bounds correspond to 2.5 and 97.5 percentiles,
 respectively. Although all ML, Ridge and LT methods almost perform identically in estimating the mixing 
 proportion and component variances, $\widehat{\bf\beta}_{ML}$  become considerably unreliable. Unlike ML 
 methods, the LT and ridge shrinkage methods could appropriately handle the multicollinearity in estimating the 
coefficients of component regressions. Comparing the shrinkage methods, we observe that the LT estimators
 appear more reliable than their ridge counterparts in estimating the parameters 
of the bone mineral population.

\section{Summary and Concluding Remarks}\label{sec:sum}
In medical and environmental research (e.g., osteoporosis research), linear regression models are used as
  standard statistical methods to investigate the relationship between a set of covariates with the response
  variable. When the underlying population is heterogeneous, the impact of the covariates on the response 
  may change in different subpopulations. A mixture of linear regression models can be considered a solution
   to the problem. The maximum likelihood (ML) method is a common technique to fit a mixture of regression 
   models; however, the ML estimates become 
unreliable in multicollinearity. We investigated shrinkage methods, including ridge and Liu-type (LT) 
estimators based on the EM, CEM and SEM algorithms in estimating the parameters of the mixture of regression 
models. We showed that LT estimators outperformed their ridge and ML counterparts through extensive numerical
 studies. Finally, we applied the developed methods to analyze the bone mineral data of women aged 50 and 
 older. While the methods were only applied to an osteoprosis research, the developed methods are generic and can be applied to
 other medical and environmental studies. 
 
\section*{Acknowledgment}  
Armin Hatefi and Hamid Usefi acknowledge the research support of the Natural Sciences and Engineering Research Council of Canada (NSERC).

\nocite{*}
\bibliographystyle{plainnat}
{\small\bibliography{LT-reg-bib}

\begin{thebibliography}{29}
\providecommand{\natexlab}[1]{#1}
\providecommand{\url}[1]{\texttt{#1}}
\expandafter\ifx\csname urlstyle\endcsname\relax
  \providecommand{\doi}[1]{doi: #1}\else
  \providecommand{\doi}{doi: \begingroup \urlstyle{rm}\Url}\fi

\bibitem[Arashi et~al.(2014)Arashi, Kibria, Norouzirad, and
  Nadarajah]{arashi2014improved}
Mohammad Arashi, BM~Golam Kibria, Mina Norouzirad, and Saralees Nadarajah.
\newblock Improved preliminary test and stein-rule liu estimators for the
  ill-conditioned elliptical linear regression model.
\newblock \emph{Journal of Multivariate Analysis}, 126:\penalty0 53--74, 2014.

\bibitem[Bliuc et~al.(2009)Bliuc, Nguyen, Milch, Nguyen, and
  Eisman]{bliuc2009mortality}
Dana Bliuc, Nguyen~D Nguyen, Vivienne~E Milch, Tuan~V Nguyen, and John~A
  Eisman.
\newblock Mortality risk associated with low-trauma osteoporotic fracture and
  subsequent fracture in men and women.
\newblock \emph{Jama}, 301\penalty0 (5):\penalty0 513--521, 2009.

\bibitem[Celeux(1985)]{celeux1985sem}
Gilles Celeux.
\newblock The sem algorithm: a probabilistic teacher algorithm derived from the
  em algorithm for the mixture problem.
\newblock \emph{Computational statistics quarterly}, 2:\penalty0 73--82, 1985.

\bibitem[Celeux and Govaert(1992)]{celeux1992classification}
Gilles Celeux and G{\'e}rard Govaert.
\newblock A classification em algorithm for clustering and two stochastic
  versions.
\newblock \emph{Computational statistics \& Data analysis}, 14\penalty0
  (3):\penalty0 315--332, 1992.

\bibitem[Cummings et~al.(1995)Cummings, Nevitt, Browner, Stone, Fox, Ensrud,
  Cauley, Black, and Vogt]{cummings1995risk}
Steven~R Cummings, Michael~C Nevitt, Warren~S Browner, Katie Stone, Kathleen~M
  Fox, Kristine~E Ensrud, Jane Cauley, Dennis Black, and Thomas~M Vogt.
\newblock Risk factors for hip fracture in white women.
\newblock \emph{New England journal of medicine}, 332\penalty0 (12):\penalty0
  767--774, 1995.

\bibitem[Dempster et~al.(1977)Dempster, Laird, and Rubin]{dempster1977maximum}
Arthur~P Dempster, Nan~M Laird, and Donald~B Rubin.
\newblock Maximum likelihood from incomplete data via the em algorithm.
\newblock \emph{Journal of the Royal Statistical Society: Series B
  (Methodological)}, 39\penalty0 (1):\penalty0 1--22, 1977.

\bibitem[Duran et~al.(2012)Duran, H{\"a}rdle, and
  Osipenko]{duran2012difference}
Esra~Akdeniz Duran, Wolfgang~Karl H{\"a}rdle, and Maria Osipenko.
\newblock Difference based ridge and liu type estimators in semiparametric
  regression models.
\newblock \emph{Journal of Multivariate Analysis}, 105\penalty0 (1):\penalty0
  164--175, 2012.

\bibitem[Faria and Soromenho(2010)]{faria2010fitting}
Susana Faria and Gilda Soromenho.
\newblock Fitting mixtures of linear regressions.
\newblock \emph{Journal of Statistical Computation and Simulation}, 80\penalty0
  (2):\penalty0 201--225, 2010.

\bibitem[Felson et~al.(1993)Felson, Zhang, Hannan, and
  Anderson]{felson1993effects}
David~T Felson, Yuqing Zhang, Marian~T Hannan, and Jennifer~J Anderson.
\newblock Effects of weight and body mass index on bone mineral density in men
  and women: the framingham study.
\newblock \emph{Journal of Bone and Mineral Research}, 8\penalty0 (5):\penalty0
  567--573, 1993.

\bibitem[Ghanem et~al.(2022)Ghanem, Hatefi, and Usefi]{elsayed_log}
Elsayed Ghanem, Armin Hatefi, and Hamid Usefi.
\newblock Liu-type shrinkage estimators for mixture of logistic regressions: An
  osteoporosis study.
\newblock \emph{Liu-type Shrinkage Estimators for Mixture of Logistic
  Regressions: An Osteoporosis Study}, Submitted:\penalty0 1--21, 2022.

\bibitem[Hatefi et~al.(2015)Hatefi, Jozani, and Ozturk]{hatefi2015mixture}
Armin Hatefi, Mohammad~Jafari Jozani, and Omer Ozturk.
\newblock Mixture model analysis of partially rank-ordered set samples: Age
  groups of fish from length-frequency data.
\newblock \emph{Scandinavian Journal of Statistics}, 42\penalty0 (3):\penalty0
  848--871, 2015.

\bibitem[Hatefi et~al.(2018)Hatefi, Reid, Jafari~Jozani, and
  Ozturk]{hatefi2018}
Armin Hatefi, Nancy Reid, Mohammad Jafari~Jozani, and Omer Ozturk.
\newblock Finite mixture modeling, classification and statistical learning with
  order statistics.
\newblock \emph{Statistica Sinica}, pages 1--50, 2018.

\bibitem[Hawkins et~al.(2001)Hawkins, Allen, and
  Stromberg]{hawkins2001determining}
Dollena~S Hawkins, David~M Allen, and Arnold~J Stromberg.
\newblock Determining the number of components in mixtures of linear models.
\newblock \emph{Computational Statistics \& Data Analysis}, 38\penalty0
  (1):\penalty0 15--48, 2001.

\bibitem[Hoerl and Kennard(1970)]{hoerl1970ridge}
Arthur~E Hoerl and Robert~W Kennard.
\newblock Ridge regression: Biased estimation for nonorthogonal problems.
\newblock \emph{Technometrics}, 12\penalty0 (1):\penalty0 55--67, 1970.

\bibitem[Hoerl et~al.(1975)Hoerl, Kannard, and Baldwin]{hoerl1975ridge}
Arthur~E Hoerl, Robert~W Kannard, and Kent~F Baldwin.
\newblock Ridge regression: some simulations.
\newblock \emph{Communications in Statistics-Theory and Methods}, 4\penalty0
  (2):\penalty0 105--123, 1975.

\bibitem[Inan and Erdogan(2013)]{inan2013liu}
Deniz Inan and Birsen~E Erdogan.
\newblock Liu-type logistic estimator.
\newblock \emph{Communications in Statistics-Simulation and Computation},
  42\penalty0 (7):\penalty0 1578--1586, 2013.

\bibitem[Jones and McLachlan(1992)]{jones1992fitting}
PN~Jones and Geoffrey~J McLachlan.
\newblock Fitting finite mixture models in a regression context.
\newblock \emph{Australian Journal of Statistics}, 34\penalty0 (2):\penalty0
  233--240, 1992.

\bibitem[Kim et~al.(2012)Kim, Yang, Cho, and Park]{kim2012relationship}
Sang~Jun Kim, Won-Gyu Yang, Eun Cho, and Eun-Cheol Park.
\newblock Relationship between weight, body mass index and bone mineral density
  of lumbar spine in women.
\newblock \emph{Journal of bone metabolism}, 19\penalty0 (2):\penalty0 95--102,
  2012.

\bibitem[Lim et~al.(2016)Lim, Kim, Lee, Byun, Park, and Kim]{lim2016comparison}
Hee-Sook Lim, Soon-Kyung Kim, Hae-Hyeog Lee, Dong~Won Byun, Yoon-Hyung Park,
  and Tae-Hee Kim.
\newblock Comparison in adherence to osteoporosis guidelines according to bone
  health status in korean adult.
\newblock \emph{Journal of bone metabolism}, 23\penalty0 (3):\penalty0
  143--148, 2016.

\bibitem[Liu(2003)]{liu2003using}
Kejian Liu.
\newblock Using liu-type estimator to combat collinearity.
\newblock \emph{Communications in Statistics-Theory and Methods}, 32\penalty0
  (5):\penalty0 1009--1020, 2003.

\bibitem[McLachlan et~al.(2019)McLachlan, Lee, and
  Rathnayake]{mclachlan2019finite}
Geoffrey~J McLachlan, Sharon~X Lee, and Suren~I Rathnayake.
\newblock Finite mixture models.
\newblock \emph{Annual review of statistics and its application}, 6:\penalty0
  355--378, 2019.

\bibitem[Melton~III et~al.(1998)Melton~III, Atkinson, O'connor, O'fallon, and
  Riggs]{melton1998bone}
L~Joseph Melton~III, Elizabeth~J Atkinson, Michael~K O'connor, W~Michael
  O'fallon, and B~Lawrence Riggs.
\newblock Bone density and fracture risk in men.
\newblock \emph{Journal of Bone and Mineral Research}, 13\penalty0
  (12):\penalty0 1915--1923, 1998.

\bibitem[Neuburger et~al.(2015)Neuburger, Currie, Wakeman, Tsang, Plant,
  De~Stavola, Cromwell, and van~der Meulen]{neuburger2015impact}
Jenny Neuburger, Colin Currie, Robert Wakeman, Carmen Tsang, Fay Plant, Bianca
  De~Stavola, David~A Cromwell, and Jan van~der Meulen.
\newblock The impact of a national clinician-led audit initiative on care and
  mortality after hip fracture in england: an external evaluation using time
  trends in non-audit data.
\newblock \emph{Medical care}, 53\penalty0 (8):\penalty0 686, 2015.

\bibitem[Pearce and Hatefi(2021)]{pearce2021multiple}
Andrew~David Pearce and Armin Hatefi.
\newblock Multiple observers ranked set samples for shrinkage estimators.
\newblock \emph{arXiv preprint arXiv:2110.07851}, 2021.

\bibitem[Peel and MacLahlan(2000)]{peel2000finite}
DAVID Peel and G~MacLahlan.
\newblock Finite mixture models.
\newblock \emph{John \& Sons}, 2000.

\bibitem[Quandt and Ramsey(1978)]{quandt1978estimating}
Richard~E Quandt and James~B Ramsey.
\newblock Estimating mixtures of normal distributions and switching
  regressions.
\newblock \emph{Journal of the American statistical Association}, 73\penalty0
  (364):\penalty0 730--738, 1978.

\bibitem[Wedel et~al.(1998)Wedel, Ter~Hofstede, and
  Steenkamp]{wedel1998mixture}
Michel Wedel, Frenkel Ter~Hofstede, and Jan-Benedict~EM Steenkamp.
\newblock Mixture model analysis of complex samples.
\newblock \emph{Journal of Classification}, 15\penalty0 (2):\penalty0 225--244,
  1998.

\bibitem[WHO(1994)]{world1994assessment}
WHO.
\newblock Assessment of fracture risk and its application to screening for
  postmenopausal osteoporosis: report of a who study group [meeting held in
  rome from 22 to 25 june 1992].
\newblock 1994.

\bibitem[Zhang et~al.(2006)Zhang, Li, and Yuen]{zhang2006mixture}
Zhiqiang Zhang, Wai~Keung Li, and Kam~Chuen Yuen.
\newblock On a mixture garch time-series model.
\newblock \emph{Journal of Time Series Analysis}, 27\penalty0 (4):\penalty0
  577--597, 2006.

\end{thebibliography}
}

\newpage
\section{Appendix}\label{app}

\subsection{ Proof of Lemma \ref{can_em_ridge}}
The  positive eigenvalues of ${\bf X}^\top {\bf W}_j {\bf X}$ and ${\bf W}_j^{1/2} {\bf X} {\bf X}^\top {\bf W}_j^{1/2}$ must be the same. Hence, 
the eigenvalue of ${\bf W}_j^{1/2} {\bf X} {\bf X}^\top {\bf W}_j^{1/2}$ are given by $\lambda_{1j},\ldots,\lambda_{pj}$ 
and the other $(n-p)$ values must be zero. From singular value decomposition, it s easy to see 
${\bf V}_1 = {\bf W}_j^{1/2} {\bf X} {\bf U}_j {\bl}_j^{-1/2}$ and   
${\bl}_j^{1/2}= {\bf V}_1^\top {\bf W}_j^{1/2} {\bf X} {\bf U}_j$. From the definition of ${\bf V}_1$ and ${\bl}_j^{1/2}$, we can show
\begin{align} \label{wjx_em_ridge}
{\bf W}_j^{1/2} {\bf X} = {\bf V}_1 {\bf V}_1^\top {\bf W}_j^{1/2} {\bf X} {\bf U}_j {\bf U}_j^\top = 
{\bf V}_1  {\bl}_j^{1/2} {\bf U}_j^\top.
\end{align}
From \eqref{wjx_em_ridge}, we can write the canonical form of the regression by
\begin{align} \label{wjy_em_ridge} 
{\bf W}_j^{1/2}  {\bf y} 
=  {\bf W}_j^{1/2} {\bf X}  {\bb}_{j} + {\bf W}_j^{1/2} \epsilon 
= {\bf V}_1  {\bl}_j^{1/2} {\bf U}_j^\top  {\bb}_{j} +  {\bf W}_j^{1/2} \epsilon 
= {\bf V}_1  {\bl}_j^{1/2} {\ba}_{j} +  {\bf W}_j^{1/2} \epsilon. 
\end{align}
From \eqref{wjy_em_ridge}, we can derive the canonical form of the weighted ridge estimator in each component by
\begin{align*} 
{\widehat\ba}_{R,j} 
&= \left( ({\bf V}_1 {\bl}_j^{1/2} )^\top  ({\bf V}_1 {\bl}_j^{1/2} ) + k_j \I \right)^{-1} 
({\bf V}_1 {\bl}_j^{1/2} )^\top  {\bf W}_j^{1/2} {\bf y} \\
&= \left( {\bl}_j^{1/2} {\bf V}_1^\top {\bf V}_1 {\bl}_j^{1/2} + k_j \I \right)^{-1}   
{\bl}_j^{1/2} {\bf V}_1^\top  {\bf W}_j^{1/2} {\bf y} \\
&= \left( {\bl}_j + k_j \I \right)^{-1}   
{\bl}_j^{1/2} {\bf V}_1^\top  {\bf W}_j^{1/2} {\bf y}.
\end{align*}
\begin{align*} 
{\bf U}_j {\widehat\ba}_{R,j} 
= & {\bf U}_j \left( {\bl}_j + k_j \I \right)^{-1}   
{\bf U}_j^\top {\bf U}_j {\bl}_j^{1/2} {\bf V}_1^\top  {\bf W}_j^{1/2} {\bf y}\\
= &  \left( {\bf U}_j {\bl}_j^{1/2} {\bl}_j^{1/2} {\bf U}_j^\top + k_j \I \right)^{-1}   
( {\bf V}_1 {\bl}_j^{1/2} {\bf U}_j^\top)^\top {\bf W}_j^{1/2} {\bf y}\\
= &  \left( ({\bf V}_1 {\bl}_j^{1/2} {\bf U}_j^\top)^\top ({\bf V}_1 {\bl}_j^{1/2} {\bf U}_j^\top) + k_j \I \right)^{-1}   
( {\bf V}_1 {\bl}_j^{1/2} {\bf U}_j^\top)^\top {\bf W}_j^{1/2} {\bf y}\\
= &  \left( ({\bf W}_j^{1/2} {\bf X})^\top ({\bf W}_j^{1/2} {\bf X}) + k_j \I \right)^{-1}   
( {\bf W}_j^{1/2} {\bf X})^\top {\bf W}_j^{1/2} {\bf y}\\
= &  \left(  {\bf X}^\top {\bf W}_j {\bf X} + k_j \I \right)^{-1}   
{\bf X}^\top {\bf W}_j {\bf y}
\end{align*}
\hfill $\square$

\subsection{ Proof of Lemma \ref{can_cem_ridge}}
The lemma can be proved in a similar  vein to Lemma \ref{can_em_ridge}. \hfill $\square$

\subsection{ Proof of Lemma \ref{can_em_lt}}
From Lemma \ref{can_em_ridge}, \eqref{wjx_em_ridge} and \eqref{wjy_em_ridge}, we can write the canonical form of the weighted LT estimator in each component regression as
\begin{align*} 
{\widehat\ba}_{LT,j} 
&= \left( ({\bf V}_1 {\bl}_j^{1/2} )^\top  ({\bf V}_1 {\bl}_j^{1/2} ) + k_j \I \right)^{-1} 
\left( ({\bf V}_1 {\bl}_j^{1/2} )^\top  {\bf W}_j^{1/2} {\bf y} -d_j {\widehat\ba}_{j}  \right)\\
&= \left( {\bl}_j^{1/2} {\bf V}_1^\top {\bf V}_1 {\bl}_j^{1/2} + k_j \I \right)^{-1}   
\left( {\bl}_j^{1/2} {\bf V}_1^\top  {\bf W}_j^{1/2} {\bf y} -d_j {\widehat\ba}_{j}  \right)\\
&= \left( {\bl}_j + k_j \I \right)^{-1}   
\left(  {\bl}_j^{1/2} {\bf V}_1^\top  {\bf W}_j^{1/2} {\bf y} -d_j {\widehat\ba}_{j}  \right).
\end{align*}
\begin{align*} 
{\bf U}_j {\widehat\ba}_{R,j} 
= & {\bf U}_j \left( {\bl}_j + k_j \I \right)^{-1}   
\left(  {\bl}_j^{1/2} {\bf V}_1^\top  {\bf W}_j^{1/2} {\bf y} -d_j {\widehat\ba}_{j}  \right)\\
= &  \left( {\bf U}_j {\bl}_j^{1/2} {\bl}_j^{1/2} {\bf U}_j^\top + k_j \I \right)^{-1}   
\left(  ( {\bf V}_1 {\bl}_j^{1/2} {\bf U}_j^\top)^\top {\bf W}_j^{1/2} {\bf y}- d_j {\bf U}_j {\widehat\ba}_{j}  \right)\\
= &  \left( ({\bf V}_1 {\bl}_j^{1/2} {\bf U}_j^\top)^\top ({\bf V}_1 {\bl}_j^{1/2} {\bf U}_j^\top) + k_j \I \right)^{-1}   
\left( ( {\bf V}_1 {\bl}_j^{1/2} {\bf U}_j^\top)^\top {\bf W}_j^{1/2} {\bf y} - d_j  {\widehat\bb}_{j}  \right)\\
= &  \left( ({\bf W}_j^{1/2} {\bf X})^\top ({\bf W}_j^{1/2} {\bf X}) + k_j \I \right)^{-1}   
\left( ( {\bf W}_j^{1/2} {\bf X})^\top {\bf W}_j^{1/2} {\bf y} - d_j  {\widehat\bb}_{j}  \right)\\
= &  \left(  {\bf X}^\top {\bf W}_j {\bf X} + k_j \I \right)^{-1}   
\left( {\bf X}^\top {\bf W}_j {\bf y} - d_j  {\widehat\bb}_{j}  \right).
\end{align*}
\hfill $\square$

\subsection{ Proof of Lemma \ref{dj_em}}
i) Since ${\widehat\ba}_{ML,j} = {\bl}_j^{-1/2} {\bf V}_1^\top  {\bf W}_j^{1/2} {\bf y} $, it is easy to show that
\begin{align} \label{al_dj_ml}
{\widehat\ba}_{LT,j} = \left( {\bl}_j + k_j \I\right)^{-1} 
({\bl}_j- d_j \I) {\widehat\ba}_{ML,j}. 
\end{align}
From \eqref{al_dj_ml}, the bias and covariance of ${\widehat\ba}_{LT,j}$ are computed by
\begin{align*}
\text{Bias} ({\widehat\ba}_{LT,j}) & = \E({\widehat\ba}_{LT,j}) - {\ba}_{j} \\
& = \left( {\bl}_j + k_j \I\right)^{-1} 
({\bl}_j- d_j \I) {\ba}_{j} - {\ba}_{j} \\
& = - \left( {\bl}_j + k_j \I\right)^{-1}
(k_j + d_j)  {\ba}_{j}. 
\end{align*} 
\begin{align*}
\text{cov} ({\widehat\ba}_{LT,j}) &= \sigma_j^2
( {\bl}_j + k_j \I)^{-1} 
({\bl}_j- d_j \I)
{\bl}_j^{-1}
({\bl}_j- d_j \I)
( {\bl}_j + k_j \I)^{-1}. 
\end{align*} 
Following \cite{liu2003using}, we then can find the $\text{MSE}({\widehat\ba}_{LT,j})$ using the bias and covariance as follows
\begin{align*}
\text{MSE}({\widehat\ba}_{LT,j}) 
&= || \text{Bias} ({\widehat\ba}_{LT,j}) ||^2 + \text{tr}(\text{cov} ({\widehat\ba}_{LT,j})) \\
&= \sum_{m=1}^{p} (d_j + k_j)^2 \alpha_m^2 / (\lambda_m + k_j)^2
+ \sigma_j^2 \sum_{m=1}^{p} (d_j - \lambda_m)^2 / \lambda_m (\lambda_m+k_j)^2.
\end{align*}
Differentiating  $\text{MSE}({\widehat\ba}_{LT,j})$ with resect to $d_j$, it is easy to obtain
\[
d_{opt,j} \sum_{m=1}^{p} \left( (\lambda_m \alpha_m^2 + \sigma_j^2 ) / \lambda_m (\lambda_m+k_j)^2\right) 
= \sum_{m=1}^{p} \left((\sigma_j^2 - k_j \alpha_m^2 )/(\lambda_m + k_j)^2  \right).
\]
ii) From Ridge method, one can easily show 
${\widehat\ba}_{R,j} = ({\bl}_j + k_j \I)^{-1} {\bl}_j^{1/2} {\bf V}_1^\top  {\bf W}_j^{1/2} {\bf y} $. Thus,
\begin{align} \label{al_dj_ridge}
{\widehat\ba}_{LT,j} = \left( {\bl}_j + k_j \I\right)^{-1} 
\left({\bl}_j + (k_j - d_j) \I\right) {\widehat\ba}_{R,j}. 
\end{align}
From \eqref{al_dj_ridge}, the bias and covariance of ${\widehat\ba}_{LT,j}$ are computed by
\begin{align*}
\text{Bias} ({\widehat\ba}_{LT,j}) & = \E({\widehat\ba}_{LT,j}) - {\ba}_{j} \\
& = \left( {\bl}_j + k_j \I\right)^{-1} ({\bl}_j + (k_j -d_j)\I) 
\left( {\bl}_j + k_j \I\right)^{-1}
 {\ba}_{j} - {\ba}_{j} \\
& = \left[ \left( {\bl}_j + k_j \I\right)^{-1} ({\bl}_j + (k_j -d_j)\I) 
\left( {\bl}_j + k_j \I\right)^{-1} -\I \right] {\ba}_{j}.
\end{align*} 
\begin{align*}
\text{cov} ({\widehat\ba}_{LT,j}) &= \sigma_j^2
( {\bl}_j + k_j \I)^{-1} 
\left({\bl}_j+(k_j- d_j) \I\right)
( {\bl}_j + k_j \I)^{-1} 
{\bl}_j
( {\bl}_j + k_j \I)^{-1} 
\left({\bl}_j+(k_j- d_j) \I\right)
( {\bl}_j + k_j \I)^{-1}.
\end{align*} 
Following \cite{liu2003using}, we then can find the $\text{MSE}({\widehat\ba}_{LT,j})$ using the bias and covariance as above. 

Similar to part (i), differentiating  $\text{MSE}({\widehat\ba}_{LT,j})$ with resect to $d_j$, it is easy to obtain
\[
d_{opt,j} \sum_{m=1}^{p} \left( \lambda_m (\sigma_j^2 + \lambda_m \alpha_m^2) / (\lambda_m+k_j)^4\right) 
= \sum_{m=1}^{p} \left(\lambda_m (\sigma_j^2 - k_j \alpha_m^2 )/(\lambda_m + k_j)^3  \right).
\]
\hfill $\square$

\subsection{ Proof of Lemma \ref{can_cem_lt}} 
The lemma can be proved in a similar  vein to Lemma \ref{can_em_lt}. \hfill $\square$

\newpage

\begin{table}
\caption{The median (M) and the length (L) of 95\% CIs for the RMSEP of the ML, ridge and LT methods 
in predicting the mixture of two regression models when $n=100$.}
\label{Pre_2Mix_size100}
\begin{centering}
\begin{tabular}{ccccccccccccccccc}
\cline{1-17}
 & $\rho$ & \multicolumn{1}{c}{} & \multicolumn{2}{c}{0.88} &  & \multicolumn{2}{c}{0.90} &  & \multicolumn{2}{c}{0.95} & 
  & \multicolumn{2}{c}{0.97} &  & \multicolumn{2}{c}{0.99}\tabularnewline
\cline{4-5} \cline{7-8} \cline{10-11} \cline{13-14} \cline{16-17} 
Method & Algorithm &  & M & L &  & M  & L &  & M & L &  & M & L &  & M & L\tabularnewline
\hline 
ML & EM &  & 16.6 & 8.2 &  & 16.9 & 8.8 &  & 17.7 & 9.0 &  & 18.1 & 8.9 &  & 18.5 & 9.3 \\[-1.5ex]
 & CEM &  & 16.5 & 8.6 &  & 16.8 & 8.7 &  & 17.6 & 9.4 &  & 18.1 & 9.0 &  & 18.5 & 9.9 \\[-1.5ex]
 & SEM &  & 16.6 & 8.5 &  & 16.9 & 8.9 &  & 17.7 & 9.0 &  & 18.0 & 9.4 &  & 18.5 & 9.8\\
\hline 
Ridge & EM &  & 16.6 & 8.8 &  & 16.8 & 8.6 &  & 17.7 & 9.1 &  & 18.1 & 9.2 &  & 18.6 & 9.4\\[-1.5ex]
 & CEM &  & 16.5 & 8.6 &  & 16.8 & 8.6 &  & 17.6 & 9.4 &  & 18.0 & 9.4 &  & 18.4 & 9.3\\[-1.5ex]
 & SEM &  & 16.5 & 8.7 &  & 16.7 & 8.5 &  & 17.8 & 9.0 &  & 18.0 & 9.3 &  & 18.5 & 9.4\\
\hline 
LT(HKP) & EM &  & 16.5 & 8.6 &  & 16.8 & 8.7 &  & 17.6 & 9.0 &  & 18.1 & 8.9 &  & 18.5 & 9.9\\[-1.5ex]
 & CEM &  & 16.4 & 8.8 &  & 16.8 & 8.5 &  & 17.6 & 8.9 &  & 18.1 & 9.4 &  & 18.4 & 9.6\\[-1.5ex]
 & SEM &  & 16.5 & 8.6 &  & 16.9 & 8.8 &  & 17.7 & 8.9 &  & 18.0 & 9.2 &  & 18.4 & 9.6\\
\hline 
LT(ITE) & EM &  & 16.6 & 8.5 &  & 16.9 & 8.4 &  & 17.5 & 9.3 &  & 17.9 & 9.3 &  & 18.2 & 9.0\\[-1.5ex]
 & CEM &  & 16.5 & 8.6 &  & 16.8 & 8.8 &  & 17.3 & 9.0 &  & 17.8 & 9.2 &  & 18.2 & 9.3\\[-1.5ex]
 & ESM &  & 16.6 & 8.4 &  & 16.8 & 8.5 &  & 17.4 & 8.8 &  & 17.8 & 9.7 &  & 18.2 & 9.5\\
\hline 
\end{tabular}
\par\end{centering}
\end{table}

\begin{table}
\caption{The median (M) and the length (L) of 95\% CIs for the RMSEP of the ML, ridge and LT methods 
in predicting the mixture of three regression models when $n=60$.}
\label{Pre_3Mix_size60}
\begin{centering}
\begin{tabular}{ccccccccccccccccc}
\hline 
 & $\rho$ & \multicolumn{1}{c}{} & \multicolumn{2}{c}{0.90} &  & \multicolumn{2}{c}{0.92} &  & \multicolumn{2}{c}{0.95} &  & \multicolumn{2}{c}{0.97} &  & \multicolumn{2}{c}{0.99}\tabularnewline
\cline{4-5} \cline{7-8} \cline{10-11} \cline{13-14} \cline{16-17} 
Method & Algorithm &  & M & L &  & M  & L &  & M & L &  & M & L &  & M & L\tabularnewline
\hline 
ML & EM &  & 6.0 & 4.0 &  & 6.1 & 3.9 &  & 6.2 & 4.1 &  & 6.4 & 4.1 &  & 6.4 & 4.3\\ [-1.5ex]
 & CEM &  & 6.1 & 3.9 &  & 6.2 & 4.0 &  & 6.4 & 4.3 &  & 6.5 & 4.3 &  & 6.6 & 4.4\\ [-1.5ex]
 & SEM &  & 6.1 & 3.9 &  & 6.2 & 4.1 &  & 6.3 & 4.2 &  & 6.4 & 4.2 &  & 6.5 & 4.3\\ 
\hline 
Ridge & EM &  & 6.0 & 4.0 &  & 6.1 & 3.9 &  & 6.2 & 4.1 &  & 6.3 & 4.2 &  & 6.4 & 4.5\\[-1.5ex]
 & CEM &  & 6.1 & 4.0 &  & 6.2 & 4.1 &  & 6.3 & 4.2 &  & 6.5 & 4.2 &  & 6.6 & 4.3\\[-1.5ex]
 & SEM &  & 6.0 & 4.0 &  & 6.2 & 4.2 &  & 6.4 & 4.1 &  & 6.5 & 4.2 &  & 6.5 & 4.3\\ 
\hline 
LT(HKP) & EM &  & 5.9 & 3.9 &  & 6.0 & 3.9 &  & 6.2 & 4.2 &  & 6.3 & 4.2 &  & 6.4 & 4.1\\ [-1.5ex]
 & CEM &  & 6.0 & 4.1 &  & 6.1 & 4.0 &  & 6.3 & 4.3 &  & 6.5 & 4.3 &  & 6.5 & 4.4\\ [-1.5ex]
 & SEM &  & 6.0 & 4.0 &  & 6.1 & 4.0 &  & 6.3 & 4.3 &  & 6.5 & 4.2 &  & 6.6 & 4.4\\ 
\hline 
LT(ITE) & EM &  & 6.5 & 6.9 &  & 6.1 & 3.9 &  & 6.1 & 4.1 &  & 6.3 & 4.0 &  & 6.3 & 4.1\\ [-1.5ex]
 & CEM &  & 6.3 & 4.4 &  & 6.3 & 4.0 &  & 6.4 & 4.2 &  & 6.5 & 4.2 &  & 6.6 & 4.3\\ [-1.5ex]
 & SEM &  & 6.2 & 4.5 &  & 6.2 & 4.0 &  & 6.3 & 4.1 &  & 6.4 & 4.2 &  & 6.5 & 4.3\\ 
\hline 
\end{tabular}
\par\end{centering}
\end{table}


\begin{table}
\caption{The median (M) and the length (L) of 95\% CIs for the RMSEP of the ML, ridge and LT methods 
in predicting the mixture of three regression models when $n=100$.}
\label{Pre_3Mix_size100}
\begin{centering}
\begin{tabular}{ccccccccccccccccc}
\hline 
 & $\rho$ & \multicolumn{1}{c}{} & \multicolumn{2}{c}{0.90} &  & \multicolumn{2}{c}{0.92} &  & \multicolumn{2}{c}{0.95} &  & \multicolumn{2}{c}{0.97} &  & \multicolumn{2}{c}{0.99}\tabularnewline
\cline{4-5} \cline{7-8} \cline{10-11} \cline{13-14} \cline{16-17} 
Method & Algorithm &  & M & L &  & M  & L &  & M & L &  & M & L &  & M & L\tabularnewline

\hline 
ML & EM &  & 5.9 & 3.0 &  & 6.1 & 3.2 &  & 6.3 & 3.1 &  & 6.4 & 3.4 &  & 6.5 & 3.4\\ [-1.5ex]
 & CEM &  & 6.1 & 3.0 &  & 6.3 & 3.1 &  & 6.4 & 3.3 &  & 6.5 & 3.3 &  & 6.6 & 3.4\\ [-1.5ex]
 & SEM &  & 6.1 & 3.0 &  & 6.1 & 3.2 &  & 6.4 & 3.3 &  & 6.5 & 3.1 &  & 6.5 & 3.4\\ 
\hline 
Ridge & EM &  & 6.0 & 3.1 &  & 6.1 & 3.1 &  & 6.3 & 3.3 &  & 6.4 & 3.2 &  & 6.5 & 3.3\\ [-1.5ex]
 & CEM &  & 6.1 & 3.0 &  & 6.2 & 3.0 &  & 6.4 & 3.2 &  & 6.5 & 3.4 &  & 6.6 & 3.4\\ [-1.5ex]
 & SEM &  & 6.0 & 3.0 &  & 6.2 & 3.1 &  & 6.4 & 3.2 &  & 6.5 & 3.3 &  & 6.6 & 3.5\\ 
\hline 
LT(HKP) & EM &  & 5.9 & 3.1 &  & 6.0 & 3.0 &  & 6.2 & 3.2 &  & 6.4 & 3.2 &  & 6.5 & 3.3\\ [-1.5ex]
 & CEM &  & 6.0 & 3.2 &  & 6.2 & 3.2 &  & 6.3 & 3.2 &  & 6.5 & 3.4 &  & 6.6 & 3.4\\ [-1.5ex]
 & SEM &  & 6.0 & 3.1 &  & 6.1 & 3.1 &  & 6.9 & 3.1 &  & 6.5 & 3.2 &  & 6.5 & 3.4\\ 
\hline 
LT(ITE) & EM &  & 6.4 & 6.2 &  & 6.1 & 3.0 &  & 6.2 & 3.1 &  & 6.3 & 3.0 &  & 6.3 & 3.5\\ [-1.5ex]
 & CEM &  & 6.3 & 3.3 &  & 6.3 & 3.2 &  & 6.4 & 3.3 &  & 6.5 & 3.4 &  & 6.6 & 3.6\\ [-1.5ex]
 & ESM &  & 6.2 & 4.2 &  & 6.2 & 3.0 &  & 6.3 & 3.1 &  & 6.3 & 3.2 &  & 6.4 & 3.4\\ 
\hline 
\end{tabular}
\par\end{centering}
\end{table}

\begin{table}
\caption{\footnotesize{The median (M), lower (L) and upper (U) bounds of 95\% CIs for $\sqrt{\text{SSE}}$ of the methods 
in the analysis of bone mineral data with sample size $n=100$.}}
\label{bone_1_size100}
\begin{centering}
\begin{tabular}{cccccccccccccc}
\hline 
 &  &  & \multicolumn{3}{c}{CEM} &  & \multicolumn{3}{c}{SEM} &  & \multicolumn{3}{c}{EM}\tabularnewline
 \cline{4-6} \cline{8-10} \cline{12-14}
Methods & $\ensuremath{\ensuremath{\boldsymbol{\Psi}}}$ &  & M & L & U &  & M  & L & U &  & M & L & U\tabularnewline
\hline 
ML & $\ensuremath{\beta}$ &  & .010 & .002 & .126 &  & .014 & .003 & .202 &  & .014 & .002 & .112\\ [-1.5ex]
 & $\ensuremath{\pi}$ &  & .350 & .100 & .380 &  & .360 & .210 & .380 &  & .222 & .016 & .370\\ [-1.5ex]
 & $\sigma^{2}$ &  & .005 & .000 & .014 &  & .006 & ..000 & .014 &  & .003 & .000 & .014\\ 
\hline 
Ridge & $\ensuremath{\beta}$ &  & .009 & .002 & .123 &  & .011 & .003 & .165 &  & .009 & .002 & .086\\ [-1.5ex] 
 & $\ensuremath{\pi}$ &  & .350 & .100 & .380 &  & .360 & .210 & .380 &  & .220 & .019 & .370\\ [-1.5ex]
 & $\sigma^{2}$ &  & .005 & .000 & .014 &  & .006 & .000 & .014 &  & .003 & .000 & .014\\ 
\hline 
LT(HKP) & $\ensuremath{\beta}$ &  & .009 & .002 & .123 &  & .010 & .003 & .133 &  & .010 & .002 & .047\\ [-1.5ex]
 & $\ensuremath{\pi}$ &  & .350 & .100 & .380 &  & .360 & .190 & .380 &  & .207 & .019 & .370\\ [-1.5ex]
 & $\sigma^{2}$ &  & .004 & .000 & .014 &  & .005 & .000 & .014 &  & .003 & .000 & .014\\ 
\hline 
LT(ITE) & $\ensuremath{\beta}$ &  & .009 & .002 & .010 &  & .009 & .007 & .010 &  & .009 & .007 & .009\\ [-1.5ex] 
 & $\ensuremath{\pi}$ &  & .310 & .100 & .380 &  & .360 & .150 & .580 &  & .584 & .040 & .600\\ [-1.5ex]
 & $\sigma^{2}$ &  & .004 & .000 & .014 &  & .005 & .000 & .014 &  & .002 & .000 & .007\\ 
\hline 
\end{tabular}
\par\end{centering}
\end{table}

\begin{table}
\caption{\footnotesize{The median (M), the lower (L) and the upper (U) bounds of 95\% CIs for the RMSEPof the ML, ridge and LT methods
 in predicting the Bone mineral population when $n=\{60,100\}$.}}
\label{MRSEP_Real}
\begin{centering}
\begin{tabular}{cccccccccc}
\hline 
 & $n$ &  & \multicolumn{3}{c}{60} &  & \multicolumn{3}{c}{100}\tabularnewline
 \cline{4-6} \cline{8-10} 
Method & Algorithm &  & M & L & U &  & M  & L & U\tabularnewline
\hline 
ML & EM &  & .139 & .105 & .195 &  & .135 & .110 & .173 \\ [-1.5ex]
 & CEM &  & .149 & .114 & .220 &  & .141 & .113 & .190\\ [-1.5ex]
 & SEM &  & .140 & .104 & .234 &  & .137 & .109 & .205\\ 
\hline 
Ridge & EM &  & .137 & .104 & .186 &  & .133 & .109 & .172\\ [-1.5ex]
 & CEM &  & .148 & .115 & .207 &  & .139 & .113 & .189\\ [-1.5ex]
 & SEM &  & .140 & .103 & .223 &  & .136 & .109 & .194\\ 
\hline 
LT(HKP) & EM &  & .135 & .104 & .182 &  & .132 & .109 & .168\\ [-1.5ex]
 & CEM &  & .148 & .115 & .208 &  & .139 & .113 & .188\\ [-1.5ex]
 & SEM &  & .139 & .104 & .221 &  & .137 & .109 & .194\\ 
\hline 
LT(ITE) & EM &  & .125 & .101 & .155 &  & .124 & .104 & .145\\ [-1.5ex]
 & CEM &  & .153 & .119 & .193 &  & .146 & .117 & .181\\ [-1.5ex]
 & SEM &  & .143 & .110 & .186 &  & .140 & .113 & .171\\ 
\hline 
\end{tabular}
\par\end{centering}
\end{table}

\end{document}